\RequirePackage{lineno} 

\documentclass[aps,prd,showpacs,twocolumn,superscriptaddress,floatfix]{revtex4}  

\usepackage{graphicx}  
\usepackage{dcolumn}   
\usepackage{bm}        
\usepackage{amssymb}   

\usepackage{natbib}

\usepackage{psfrag}

\def\pp{$p\bar{p}$}

\def\ttbar{$t\bar{t}$}

\def\lmet{$WH\rightarrow \ell\kern-0.45em\raise0.19ex\hbox{/} \nu b\bar{b}$}

\def\vhe{$VH \rightarrow e^{\pm} \nu_e \mu^{\pm} \nu_{\mu} + X$}

\def\etal{{\it et.~al.}}

\newcommand{\MET}{\ensuremath{\not\hspace*{-0.84ex}E_T\,}}

\def\MET{{\mbox{$E\kern-0.57em\raise0.19ex\hbox{/}_{T}$}}}
\def\met{{\mbox{$E\kern-0.57em\raise0.19ex\hbox{/}_{T}$}}}

\def\DZero{D0 }

\def\pp{$p\bar{p}$}

\def\ttbar{$t\bar{t}$}

\def\lmet{$WH\rightarrow \ell\kern-0.45em\raise0.19ex\hbox{/} \nu b\bar{b}$}

\def\vhe{$VH \rightarrow e^{\pm} \nu_e \mu^{\pm} \nu_{\mu} + X$}

\def\etal{{\it et.~al.}}

\def\ppbar{{p\overline{p}}}
\def\bbbar{{b\overline{b}}}

\def\ttbar{{t\overline{t}}}
\def\qqbar{{q\overline{q}}}
\def\epem{e^+e^-}

\def\tautau{{\tau^{+}\tau^{-}}}

\begin{document}




\pacs{13.85.Rm, 14.80.Bn}

\title{\boldmath  
\vspace{0.5cm}
Standard Model Higgs Boson Searches through the 125~GeV Boson Discovery}
\author{Gregorio Bernardi}

\address{LPNHE, Universit\'es Paris VI \& VII, CNRS/IN2P3, Paris, France}

\author {Matthew Herndon}

\address{University of Wisconsin, Madison, Wisconsin 53706-1390,  USA}


\begin{abstract}
\vskip 0.5cm
\noindent
Searches for the standard model Higgs boson are reviewed
from the 2 TeV run  of the Tevatron with $\simeq$ 10 fb$^{\rm -1}$ of recorded data,
 and from the 7 and 8 TeV runs of the LHC, with $\simeq$ 5 and $\simeq$ 6~fb$^{\rm -1}$,
respectively, i.e., until the July-2012 discovery of a new particle
by the LHC experiments. 
The CMS
and ATLAS Collaborations observe independently a new boson with  mass  $\simeq$ 125 GeV,
mainly through its bosonic decays in $\gamma \gamma$, $ZZ$, and $W^+W^-$, 
consistent with the standard model Higgs boson. The CDF and D0 experiments combine
their results to see evidence of a similar particle produced in association
with a vector boson and decaying fermionically in $b \bar{b}$.
\end{abstract}
\maketitle

\tableofcontents

\vspace{0.3cm}

\section{Introduction}

The origin of the masses of elementary particles, one of the remaining puzzles of the highly
successful theoretical model of the standard model (SM), has a
potential solution requiring the existence of only one doublet of complex
scalar fields. Then the finite mass of the SM elementary fermions and bosons 
can be explained, after spontaneous electroweak symmetry breaking (EWSB) of the originally massless
Lagrangian~\cite{higgs,higgs2,higgs3,sm,sm2,sm3,sm4}. This minimal approach could be confirmed if the remnant of such a
breaking, the Higgs boson, is observed with the couplings and properties predicted in the SM.
The SM Higgs boson does not solve all problems related to the EWSB, and is maybe
only one component of the fields involved.   However, its discovery would be a major step
in the final validation of the SM in that it would show
a unified approach to the generation of boson and fermion masses.
While there are other approaches to explain  EWSB, none is so far as successful
as the Higgs mechanism, so we concentrate in this review on the experimental
searches for such a boson.  
With the recent observation of a new boson at the LHC, 
a major step forward has been accomplished, but a complete validation has yet to be done.
The data taking has been concluded at the Tevatron and at the LHC for the center of mass energy of 7 TeV, 
while the data are still being accumulated at 8 TeV. However, given the crucial discovery made using these
searches, we provide a dedicated review based on the publications
which immediately followed the discovery and on the results available at that 
time, without extending the results reviewed beyond discovery time.

To review these milestone results, we first
briefly explain the phenomenology of the production and decay of the SM Higgs boson,
the indirect and direct constraints on the SM Higgs from other measurements,
and the search strategies at hadron colliders, first pioneered at the Tevatron and then
extended at the LHC.
We review the Tevatron experiments, and their low-mass and high-mass
analyses, then the LHC experiments, and their searches in bosonic and fermionic
Higgs boson decays. Finally we review the combinations of these searches, 
which led to the discovery of a new boson
by ATLAS and CMS independently, and to the evidence, from the Tevatron, 
for a particle 
consistent with this new boson.  
We 
conclude by briefly discussing the current
knowledge on this new boson and on short term prospects.

\section{SM Higgs boson phenomenology and search strategies}

	\subsection{ Phenomenology of SM Higgs Boson production}	
The SM Higgs boson is a $CP$-even scalar, and its couplings to 
fermions and to gauge bosons are proportional to the fermion masses, and to
the squares of the boson masses, respectively.
The  effective Higgs boson-gluon coupling
$Hgg$ is dominated at leading order (LO) by a one-loop graph
in which the $H$ couples dominantly to a virtual $\ttbar$ pair.
The much weaker effective coupling to photons $H \gamma \gamma$  proceeds
also at LO via a loop, dominated by a virtual $W^+W^-$ pair~\cite{hhg}.  
The dominant cross section for Higgs boson production
is the $gg \rightarrow H$ ($gg$H) process, which is known at
next-to-next-to-leading  order~(NNLO) from perturbative
calculation in quantum chromodynamics (QCD).  This calculation is
performed using a  large top-mass limit approximation and a
similar calculation at next-to-leading order (NLO) in QCD has been performed for arbitrary top mass~\cite{smxs-ggh,smxs-ggh2,smxs-ggh3,smxs-ggh4,smxs-ggh5}.  The NLO QCD
corrections approximately double the leading-order prediction, and the
NNLO corrections add approximately $50$\% to the NLO prediction.  NLO
electroweak corrections range between 0\% and 6\% of the LO term~\cite{NLO-EW,NLO-EW2,NLO-EW3}. 
Mixed QCD-electroweak corrections $O(\alpha \alpha_s)$ are also included~\cite{Anastasiou:2008tj}.
Soft-gluon contributions to the cross sections have been 
resummed at next-to-leading logarithmic (NLL),  next-to-next-to-leading logarithmic (NNLL)  and partial next-to-next-to-next-to-leading logarithmic (NNNLL) 
accuracy ~\cite{res_nnll,res_nnll2,deFlorian:1999zd,Ravindran:2002dc,res_nnll3,res_nnll4,res_nnll5,res_nnll6,res_nnll7,deFlorian:2009hc}.
Predictions for the gluon fusion cross sections at NNLO or through
soft-gluon resummation up to NNLL
accuracy  and two-loop electroweak effects 
can be found in Ref.~\cite{Anastasiou:2008tj,deFlorian:2009hc,Baglio:2010ae,Anastasiou:2012hx},
including differential cross section as a function of Higgs boson transverse momentum~\cite{deFlorian:2011xf,Bagnaschi:2011tu}. 
Uncertainties are dominated by the modeling of parton distribution functions (PDFs) and 
choices of fragmentation and renormalization scales and are $\simeq$ 25\% at the Tevatron and $\simeq$ 15\%
at the LHC.
The cross sections have also been computed exclusively for Higgs boson production in association
with one jet~\cite{Schmidt:1997wr,Glosser:2002gm}
and in association with two jets~\cite{Campbell:2006xx}\cite{Campbell:2010cz}.

\begin{figure}[b]
\vskip -0.4cm
\hspace{-1.0cm} \includegraphics[width=3.4in]{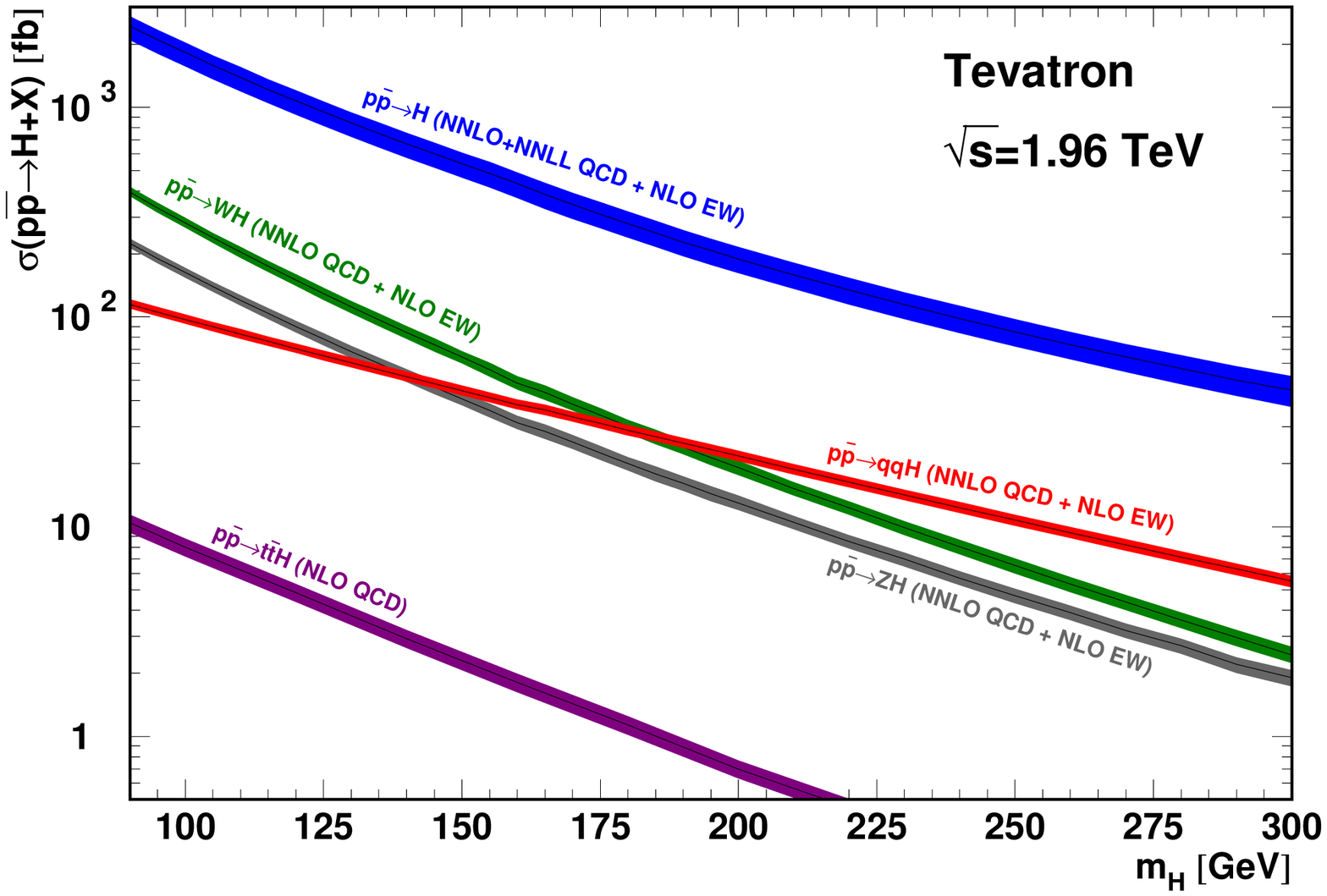}
\caption{SM Higgs boson production cross sections for $\ppbar$ collisions at 
1.96~TeV~\cite{smxs-ggh,smxs-ggh2,smxs-ggh3,smxs-ggh4,smxs-ggh5,smxs-assoc,smxs-assoc2,smxs-assoc3,smxs-assoc4,smxs-assoc5,smxs-assoc6,smxs-vbf,smxs-vbf2,smxs-vbf3,smxs-vbf4,smxs-vbf5,smxs-vbf6,smxs-vbf7,smxs-tth,smxs-tth2,smxs-tth3,smxs-tth4} as functions of
its mass.}
\label{tevxs}
\end{figure}

At the Tevatron, the next most important production processes are Higgs boson production in
association with vector bosons ($VH$), where $V$ is a massive $W$ or $Z$ vector boson.
The cross sections for $\qqbar \rightarrow WH$ or $ZH$ are
known at NNLO for the QCD corrections and at NLO for the electroweak 
corrections~\cite{higgswg2003,smxs-assoc,smxs-assoc2,smxs-assoc3,smxs-assoc4,smxs-assoc5,smxs-assoc6}, with a total uncertainty of 
$\simeq 5\%$. 
For the vector boson fusion (VBF)  process $qq \rightarrow qq H $, which dominates
over the associated production at the LHC, the
production cross sections are known at NNLO in
QCD and at NLO for the electroweak corrections, with a total theoretical 
uncertainty $\simeq$ 5\%~\cite{smxs-vbf,smxs-vbf2,smxs-vbf3,smxs-vbf4,smxs-vbf5,smxs-vbf6,smxs-vbf7}, which becomes larger when exclusive
requirements are put on the jets~\cite{lhc_differential}. 
For the associated production process $\ttbar H$, the cross section 
has been calculated at NLO in QCD~\cite{smxs-tth,smxs-tth2,smxs-tth3,smxs-tth4}.

\begin{figure}[t]
\vskip -0.4cm
\hspace{-1.0cm} \includegraphics[width=3.4in]{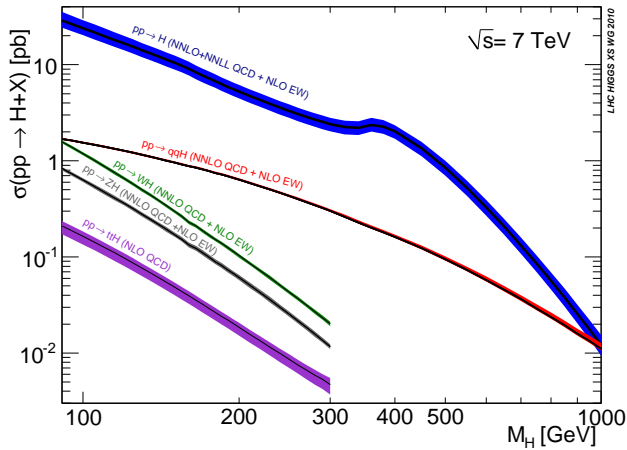}
\caption{SM Higgs boson production cross sections for $pp$ collisions at 7~TeV~\cite{lhc_inclusive},
as functions of its mass.}
\label{lhcxs}
\end{figure}

The cross sections for the production of SM Higgs bosons are
summarized in~Fig.~\ref{tevxs} for $\ppbar$ collisions at the Tevatron, and
in~Fig.~\ref{lhcxs} for $pp$ collisions at the LHC at $\sqrt{s}= 7$ TeV~\cite{smxs,lhc_inclusive}.
Cross sections at $\sqrt{s}= 8$ TeV have a similar behavior but
are 20-30\% larger at low Higgs boson mass ($m_H < 135$ GeV).

	\subsection{ Phenomenology of SM Higgs boson decay}

\begin{figure}[b]
\vskip -0.4cm
\hspace{-1.0cm} \includegraphics[width=3.4in]{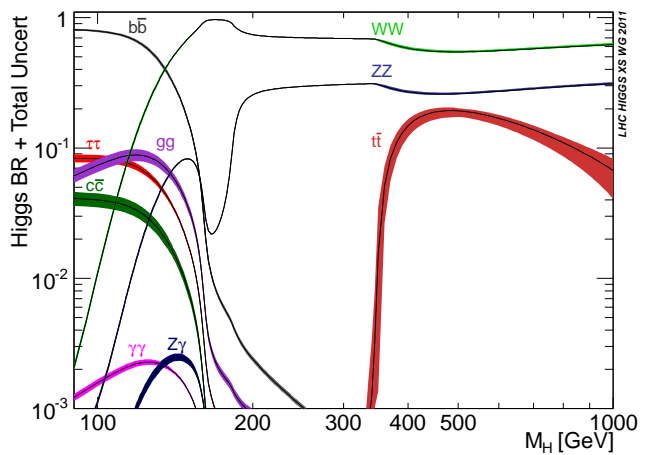}
\caption{Branching ratios for the main decays of the SM Higgs 
boson as functions of its mass}  
\label{smdecay}	
\end{figure}

The branching ratios for the most relevant decay modes of the SM 
Higgs boson are shown in~Fig.~\ref{smdecay} as functions of $m_H$. 
For masses below  135~GeV,
decays to fermion pairs dominate, of which the decay 
$H \to \bbbar$ has the largest branching ratio.  For these lower masses,
the total decay width is less than 10~MeV.
For Higgs boson masses above~135 GeV, the $WW$ decay dominates 
with an important contribution from $H \to ZZ$ above threshold.
The decay width rises rapidly,
reaching about 1~GeV at $m_H = 200$~GeV and 100~GeV at 
$m_H = 500$~GeV.
Above the $\ttbar$ threshold, the branching ratio into top-quark pairs
increases rapidly as a function of the Higgs boson mass, reaching a maximum of
about~20\% at $m_H \simeq 450$~GeV.

	\subsection{Standard model fits}
\begin{figure}[tbh]
\vskip -0.4cm
\hspace{-1.0cm} \includegraphics[width=3.4in]{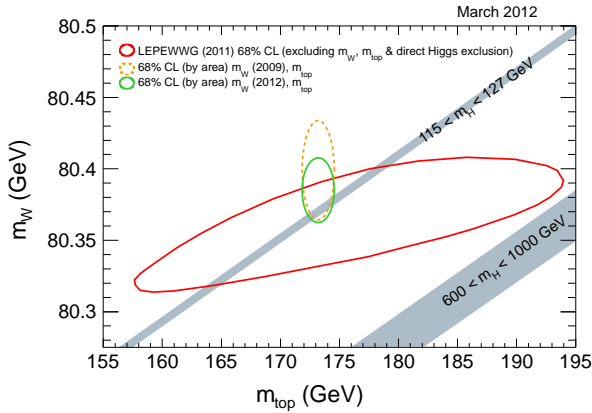}
\caption{Constraints from the top-quark mass, the $W$ boson mass and other SM measurements compared to the allowed
Higgs search range in March 2012.}
\label{mw_vs_mtop}
\end{figure}

While the mass of the SM Higgs boson is not given by the theory, 
indirect constraints for the SM Higgs boson mass can be derived
from fits to precision measurements of electroweak observables. The
Higgs boson contributes to the observed $W$ and $Z$ masses through
loop effects, leading to a logarithmic sensitivity of the ratio of 
the $W$ and $Z$ gauge boson masses on the Higgs boson mass.
The top quark contributes to the observed $W$ boson mass through
loop effects that depend quadratically on the top mass, which thus also plays 
an important role in the global fit.  
A global fit to precision electroweak
data, accumulated over the last two decades mainly at LEP, SLC and the Tevatron, 
gives $m_H = 94^{+29}_{-24}$~GeV   
or $m_H < 152$~GeV at 95\%~C.L.~\cite{elweak,elweaknp}.
Measurements of the top-quark mass  
[$173.2\pm0.9$~GeV~(\cite{newtopmass})]
and of the $W$ boson mass [$80.385\pm 0.015$~GeV~(\cite{tev-w-mass})]
were used for these constraints.
These results,  compared to the allowed direct search range from March 2012
are shown in Fig.~\ref{mw_vs_mtop}.

	\subsection{Direct constraints from LEP}

At the LEP $e^+e^-$ collider, which operated between 1989 and 2000 at $\sqrt{s}=90$ GeV 
(LEP1) or $160$-$209$ GeV (LEP2),
the SM Higgs boson is generally produced 
through Higgsstrahlung in the $s$ channel,$\epem\rightarrow HZ$~\cite{bjorken,ellis1976},
where the $Z$ boson in the final state is either virtual (LEP1) or
on shell (LEP2).  
The SM Higgs boson can also be produced by
$WW$ and $ZZ$ fusion in the $t$ channel~\cite{fusion,fusion2,fusion3},
but at LEP these processes have small cross sections.
The sensitivity of the LEP searches to the Higgs boson 
strongly depends on the center of mass energy $E_{\rm{cm}}.$  
For $m_H < E_{\rm{cm}}-m_Z$, the cross section
is of the order of 1~pb or more, while for $m_H > E_{\rm{cm}}-m_Z$, 
the cross section is smaller by at least an order of magnitude.

Each production and decay mode was analyzed separately.
Data recorded at each center of mass energy were studied independently
and the results from the four LEP experiments were then combined.  
Strong upper bounds on the $e^+e^-\to ZH$ cross section
are obtained for Higgs boson masses between $1$~keV and $\simeq$ 115~GeV.
The combination of the LEP data yields a 95\%~C.L. lower bound of~114.4~GeV for the mass
of the SM Higgs boson~\cite{sm-lep}.  The median limit expected in a large
ensemble of identical experiments when no signal is present (simply called
``expected limit'' in the following) is 115.3~GeV  and was limited
by the collision energy achieved by the accelerator.  

	\subsection{Search strategies at the Tevatron}

At the Tevatron, the delivered luminosity and the Higgs boson production cross section
are sufficient to be sensitive at the 95\% C.L. to a Higgs boson having a mass between
90 and 185 GeV, i.e., significant overlap with the LEP excluded region at lower
masses, and with the LHC. Sensitivity is strongest at lower mass,
90-120 GeV and around the $H \to WW$ threshold 
($\simeq 145-185$ GeV) where the best expected sensitivity reaches approximately 4 standard deviations.
The search strategy gives priority to direct production and the $H \to WW$ decay mode at high mass,
which allowed in 2009 for the extension of the experimental limits on
the SM Higgs boson mass beyond the LEP exclusion limit for the first time after almost ten years~\cite{Tev-162-166},
with subsequent enlarged excluded regions~\cite{tev-2010,tev-2011}.
These results in combination with the precision electroweak measurements showed,
before the LHC produced significant results, that a mass 
of the SM Higgs boson above $\simeq 145$ GeV was excluded at 95\%C.L.~\cite{gfitter}

At low-mass, the associated production channels ($WH, ZH$) 
involving $H \to b\bar{b}$ are the most sensitive, 
with a significant additional contribution from direct production with $H \to WW$
decay at Higgs boson masses as low as 120 GeV. The
$WH$ and $ZH$ channels with $H \to b\bar{b}$ are particularly important 
since this decay mode will probably not be observed at the $5 \sigma$ level at ATLAS or
CMS before additional statistics has been accumulated after the upgrade 
to the full energy foreseen for 2015. Even though it will be measured more precisely in the
LHC experiments
than at the Tevatron, once the full 2011-2012 statistics will have been
analyzed, the contribution of the Tevatron results will still be significant.

	\subsection{Search strategies at the LHC}

The LHC was designed to have a full reach to discover the Higgs boson, from 0.1 to 1 TeV, 
which is the region where it was theoretically expected to be. 
The high energy of the LHC proton-proton collider substantially increases
the cross section for production of a Higgs boson via gluon fusion ($ggH$).  The
cross section for vector boson fusion and associated production are also 
enhanced.  Because of a more unfavorable signal-to-background ratio in associated production (due to
the initial dominant $qg$ or $gg$ vs $q\bar{q}$ hard interaction),
the search strategy gives priority to inclusive production and the $H \to \gamma\gamma$ or $H \to ZZ$ decay modes at lower masses,
or simply to the $H \to ZZ$ decay mode at higher masses. 
Both modes allow possible discovery 
as soon as the available integrated luminosity
is sufficient (of the order of 10 fb$^{\rm -1}$).
Also at high mass $ZZ$ and $W^+W^-$ Higgs boson decay modes with  $W$ and $Z$ decays to jets
or $Z$ decays to neutrinos have strong sensitivity. 
The $H \to WW$ channel is best searched in the region around the $WW$ on-shell decay threshold, but does
not have the mass resolution to observe a clear resonance signal.
No single channel dominates the sensitivity,
so the search is also performed through combination of all channels.
Already as of 2011 this strategy has allowed for the exclusion of a Higgs
boson between masses of approximately 130 to 500 GeV.

In addition, once a new particle is observed all channels are crucial to understanding its nature.
In particular the fermionic decays (in pairs of $\tau$ leptons or $b$ quarks) are a major source
of information on fermion couplings, but may require a higher luminosity. 
In both channels it is possible to find evidence for fermionic
decays and measure Yukawa couplings of the Higgs boson to the fermions given
SM couplings using data collected before the upgrade to the full energy foreseen for 2015. 
The $\tau$ mode is the stronger channel since it can be seen in
all Higgs boson production mechanisms with manageable signal to background.
Other fermionic decays will not be not observable
before a significant luminosity upgrade not foreseen before several years, 
assuming SM branching ratios, although information is
gained about the Higgs boson coupling to the top quark through the loop production diagrams.

\subsection{Simulation of background and signal processes}

The general strategies used to generate background and signal
processes at both the Tevatron and LHC are summarized in this section.

{i) High cross section backgrounds $W/Z$ + jets:}
 the SM background processes $Wq\bar{q} \rightarrow 
\ell \nu q\bar{q}$ or $Zq\bar{q}\rightarrow \ell^+\ell^- q\bar{q}$, 
where $q$ is used to represent light partons $u,d,s$ and gluons ($g$), 
and higher cross section diboson processes,
are simulated using Monte Carlo (MC) matrix element event generators such as
{\tt ALPGEN}\cite{Mangano:2006rw}, {\tt MADGRAPH}\cite{madgraph}, and {\tt POWHEG}\cite{powheg,powheg2,powheg3}. 
Separate samples are generated for light parton 
multiplicities and in each case samples are generated
for each of the final state decay lepton flavors $\ell=e,\mu,\tau$.
To account for the subsequent hadronization and development of partonic 
showers, the matrix event generators are interfaced to {\tt PYTHIA}\cite{pythia} 
using the M.L. Mangano (MLM) factorization (``matching'') scheme~\cite{Mangano:2006rw} to remove events
where soft jets developed in partonic showering overlap phase space
already covered by the matrix event generator.

{ii) top-antitop, ``single'' top and diboson production:}
$\ttbar$ production has been studied since the Tevatron run 1 and 
is now also studied at the LHC.
The electroweak
production of (single) top-quark has been observed at the Tevatron by
the CDF and D0 collaboration in 2008~\cite{cdf-singletop,d0-singletop}
and also more recently by the LHC experiments.
Similarly diboson production ($WW, WZ, ZZ$) has been measured at the Tevatron and LHC. 
Extensive studies of top, single top, and diboson production have been
performed by the collaborations
showing that the normalization and kinematic distributions of these processes are well modeled
by a variety of programs including  {\tt
  CompHep}\cite{comphep,comphep2}, {\tt MC@NL0}\cite{MC@NLO}, {\tt MADGRAPH}, and {\tt POWHEG} 
interfaced to {\tt PYTHIA}. 
The diboson processes $WW$ and $ZZ$ also have substantial contributions from gluon-gluon
initial states which are generated using the specialized 
generators {\tt GG2WW}\cite{gg2ww} and {\tt GG2ZZ}\cite{gg2zz}
at LHC experiments.
In cases where there are multiple additional partons in the final state, the data are not
as constraining and the techniques listed in section i) are applied.

{iii)  backgrounds with $b$-flavored quarks, $Wb\bar{b}$, $Zb\bar{b}$:}
These production processes are generated using similar techniques to those
described above.  
Normalization k-factors measured
in data are applied 
to these associated vector boson with $b$-flavored quark production processes,
as their total cross sections are not precisely predicted.
Scale factors to account for efficiency
differences in simulation and data for $b$-jet identification are applied,
as is also the case for other processes with $b$-jet in the final state.

iv) Higgs boson signal samples: these are generated using 
programs such as {\tt PYTHIA} and {\tt POWHEG} interfaced
to {\tt PYTHIA}.  In the case of gluon fusion production the kinematics calculated at NNLO
can differ from leading order generation and samples are re-weighed to differential NNLO
calculations.

For each background or signal the total cross section is normalized to the best available NLO or NNLO
calculations.

\section{The Tevatron and  the CDF and D0 detectors}

The Tevatron was a proton-antiproton collider which completed operation in September 2011.  The Higgs boson searches took place during
run II (2002-2011) in which 
it was configured to collide beams of 36 bunches with 1.96 TeV center of mass energy and provided
an integrated dataset of  10~fb$^{\rm -1}$ to the CDF and D0 experiments (note
integrated luminosities given in this review refer to integrated
luminosity delivered with the detectors in an operational
condition sufficient to be used in physics analysis).
The instantaneous luminosity reached 4x10$^{\rm 32}$ cm$^{\rm 2}$ s$^{\rm -1}$, but
the effect of the overlay of multiple interactions remained manageable.

The main components of the run II CDF and D0 detectors
are the tracking detectors, calorimeters and  muon detectors.
Specific details 
of the  CDF and D0 detector subsystems are available in 
~\cite{Abulencia:2005ix} and ~\cite{d0det_run2}, respectively, while
here we briefly summarize their main characteristics.
The kinematic properties of
particles and jets are defined with respect to the origin of the detector
coordinate system which is at the center of the detector. 
To quantify polar angles the pseudorapidity  variable, 
defined as $\eta = - \ln \tan (\theta/2$), is used  where $\theta$ is the polar angle 
in the corresponding spherical polar coordinate system.

\subsection{Tracking detectors}

The CDF tracking system consists of an eight layer silicon micro-strip tracker
and an open-cell drift chamber referred to as the central outer tracker (COT),
both immersed in a 1.4~T solenoidal magnetic field.  These systems combined
provide charged particle tracking and precision vertex reconstruction in
the pseudorapidity region $| \eta | <1.0$ with partial coverage in the COT
to $| \eta | < 1.7$ while the two outer layers
of the silicon detector extend the
tracking capability to $| \eta | < 2.0$.  
 
The D0 tracking system is located immediately 
surrounding the interaction point and consists of an inner silicon
micro-strip tracker (SMT) surrounded by an outer central scintillating 
fiber tracker (CFT). Both the SMT and CFT are situated within a 2 T 
magnetic field provided by a solenoidal magnet surrounding the entire
tracking system.
The SMT is used for tracking up to $|\eta| < 2.5$
and for vertex 
reconstruction. 
The central fiber tracker  is also used for tracking and vertex reconstruction,
and provides precise tracking coverage up to $|\eta| < 1.7$.

\subsection{Calorimeters}

The CDF calorimeter systems are used to measure the energy of charged and 
neutral particles produced in $p\bar{p}$ collisions and are arranged around the 
outer edges of the central tracking volume and solenoid. These systems consist of modular 
sampling scintillator calorimeters with a tower based projective geometry. 
The inner electromagnetic sections of each tower consist of lead sheets 
interspersed with scintillator, and the outer hadronic sections are composed 
of scintillator sandwiched between sheets of steel. The CDF calorimeter 
consists of two sections: a central barrel calorimeter and forward end plug 
calorimeters covering the pseudorapidity region $ |\eta| <3.64$.  The 
calorimeters can identify and measure photons, jets from partons, missing 
transverse energy, and in combination with information from other systems electron 
and tau leptons.

The D0 liquid-argon calorimeter system is used for the identification and energy measurement of electrons, 
photons, and jets, and also allows the measurement of the missing transverse energy (\MET )
of the events, typically from unobserved neutrinos. 
The central calorimeter (CC) covers detector pseudorapidities $|\eta| \le 1.1$ 
and the two additional end-cap calorimeters extend the range up to $|\eta| = 4.2$. 
They are located outside of the tracking and solenoid systems.
The calorimeters are subdivided into electromagnetic (EM) followed by fine hadronic and 
then coarse hadronic sections. 
The intercryostat plastic scintillator detectors 
complete the calorimeter coverage in the intermediate pseudorapidity region $0.8<|\eta|<1.4$.

\subsection{Muon detectors}

The CDF muon detector is made up of four independent detector systems outside 
the calorimeter modules and consists of drift chambers interspersed with
steel layers to absorb hadrons. The central muon detector (CMU) is mounted 
directly around the outer edge of the central calorimeter module and detects
muons in the  pseudorapidity region $| \eta | <0.6$.  The central muon 
extension is composed of spherical sections and extends the pseudorapidity 
coverage  in the range $0.6< | \eta | <1.0$.  The central muon upgrade (CMP) surrounds portions of the CMU 
and central muon extension (CMX) systems covering gaps in angular coverage and allowing excellent 
identification of higher
momentum muons due to additional layers of steel absorber.  
The barrel muon upgrade (BMU)
is a barrel shaped extension of the muon system in the pseudorapidity 
region $1.0< | \eta | <1.5$.  The CMX,
CMP and BMU systems also include
matching scintillator systems which provide timing information to help
identify collision produced muons.

The D0 muon detector system consists of a central muon detector system 
covering the range $|\eta|<1$ and a forward muon system which covers
the region $1<|\eta|<2$. 
Scintillation counters are included for triggering purposes
and a 1.8 T toroidal iron magnet makes it possible to determine muon momenta and 
perform tracking measurements within the muon system alone, although in general the
central tracking information is also used for muon reconstruction.

\subsection{Triggering systems}

The CDF trigger system consists of three levels.  Level one trigger hardware consists
of dedicated electronics that operate at the beam crossing frequency. 
The level one trigger can identify and measure the transverse momentum of
charged particles using COT information
and be combined with information from the calorimeters or
muon systems to provide a trigger for leptons.  The calorimeter
trigger hardware measures energy clusters which are used to identify  jets and photons
as well as an imbalance in event transverse energy interpreted as \MET.  The
second level trigger hardware at CDF refines the measurements of the level
one trigger at higher precision.  The level two trigger can also include
tracking and vertexing information from the silicon detectors.   The third
level of the trigger operates on commercial computers (PCs) and
executes fast versions of the full offline reconstruction software.

The D0 trigger system also has three 
trigger levels referred to as L1, L2 and L3. Each consecutive
level receives a lower rate of events for further
examination. 
The L1 hardware based elements of the 
triggers used in the electron channel typically require calorimeter
energy signatures consistent with an electron. This is  expanded at L2
and L3 to 
include trigger algorithms requiring an electromagnetic object together with at least 
one jet for which the L1 requirement is calorimeter energy depositions consistent with 
high-$p_{T}$ jets. 
For muon samples, events are triggered using
the logical .OR. of the full list of available triggers
of the D0 experiment. The muon trigger pseudorapidity coverage is restricted
to  $|\eta| < 1.6$ where the majority of the $W$+jet events ($\simeq$65\%) are 
collected by triggers requiring high-$p_{T}$ muons at L1. 
Events not selected by the high-$p_{T}$ muon triggers 
are primarily collected by jet triggers.

\subsection{Physics object identification at the Tevatron}

\subsubsection{Lepton identification}

Isolated electrons are reconstructed in the calorimeter
and are selected 
in the pseudorapidity regions 
$|\eta|<2.8$ at CDF, and at
$|\eta|<1.1$ and $1.5 < |\eta| < 2.5$ at D0. 
The EM showers are required to pass spatial distribution requirements consistent
with those expected from electrons for each section of the calorimeter. In the D0 CC region, a 
reconstructed track, isolated from other tracks, is also required to be matched 
to the EM shower while in CDF a matching track is required within the coverage of the COT tracker.

Muons are selected by requiring 
a local track spanning all layers of the muon detector system (for D0 both within as well as outside of the
toroidal magnet).  A spatial match is then required to a corresponding track 
in the COT (CDF) or CFT (D0). 
To suppress muon background events originating from the semileptonic 
decay of hadrons, muon candidate tracks are required to be separated from
jets by at least $\Delta R = \sqrt{(\Delta \eta)^{2} + (\Delta \varphi)^{2}} > 0.5 $. 
A veto against cosmic ray muons is also applied using scintillator timing information in D0 and
a specialized tracking algorithm is used at CDF to track cosmic ray muons 
passing through both sides of the detector.
Muons can also be identified as a minimum ionizing isolated track in CDF for regions without
muon coverage taking advantage of the fact that muons interact with
low probability in the material of the calorimeter leaving only a
small ionization signature..

Multivariate algorithms (MVA)~\cite{mva}
are used to enhance efficiency and background rejection
in some electron and muon based analysis.

Tau lepton decays into hadrons are characterized as narrow, isolated jets with 
lower track multiplicities than jets originating from quarks and gluons. Three types of tau lepton 
decays are distinguished by their detector signature. One-track tau decays 
consisting of energy deposited in the hadronic calorimeter associated with a 
single track are denoted as tau-type 1; tau-type 2 corresponds to one-track tau 
decays with energy deposited in both the hadronic and EM calorimeters, 
associated with a single track; and tau-type 3 are multitrack decays with 
energy in the calorimeter and two or more associated tracks with invariant mass 
below 1.7 GeV.  
In D0, a set of neural networks, one for each tau type, is applied 
to discriminate hadronic tau decays from jets. The input variables are 
related to isolation and shower shapes, and exploit correlations between 
calorimeter energy deposits and tracks. When requiring the neural network 
discriminants to be above thresholds optimized for each tau type separately,
typically  65\% of taus are retained, while 98\% of the multijet (MJ) background is 
rejected.  In CDF boosted decision tree (BDT) based algorithms are used for the same purpose.

\subsubsection{Jets, $b$ jets, and missing transverse energy}

In CDF, jets are reconstructed using a calorimeter based clustering algorithm,
with a cone of size $\Delta R < 0.4$.
In D0, jets are reconstructed in the calorimeters
for  $|\eta| < 2.5$ using the D0 run II iterative cone algorithm.
Calorimeter energy deposits 
within a cone of size $\Delta R < 0.5 $ are used to form the jets.
The energy of the jets is calibrated by applying 
a jet energy scale correction determined using $\gamma$+jet events.

Jet identification  efficiency and jet resolutions are adjusted in the simulation to match those measured in data.
At high instantaneous luminosity, the jets are further required
to contain at least two 
tracks with $p_{T}>0.5 ~\rm GeV$ 
associated with the primary vertex at D0.

At both CDF and D0, the identification of quarks initiated by
a $b$ quark (``$b$ tagging'') is done in two steps~\cite{d0-btag}.
The jets are first required to pass a taggability requirement based on charged particle 
tracking and vertexing information, to ensure that they originate from the interaction vertex and
that they contain charged tracks.
At D0  a $b$ tagging NN is 
applied to the taggable jets. This NN uses a combination 
of seven input variables, five of which contain secondary vertex information; the 
number and mass of vertices, the number of and $\chi^{2}$ of the vertex contributing
tracks, and the decay length significance in the $x-y$ plane. Two impact parameter 
based variables are also used. 
At CDF the next step in $b$ tagging is done using a NN with similar variables but including additional track
quality information~\cite{hobit}.  The CDF experiment also employs a cut based secondary
vertex tagger~\cite{Acosta:2005am}.
As an example at D0 the typical efficiency for identifying a $p_T = 50~\rm GeV$ jet 
that contains a $b$ hadron is $(59\pm 1)$\% at a corresponding misidentification rate of 1.5\%\ for light 
parton ($u,d,s,g$) initiated jets. This operating point is typically used for events with two
``loose'' ($L$) $b$-tagged jets.
When tightening ($T$) the identification requirement, the efficiency for identifying a jet
with $p_T$ of 50~GeV 
that contains a $b$ hadron is $(48\pm 1)$\% with a misidentification rate of 0.5\%\ for 
light parton jets.
The event missing transverse energy (\MET )is calculated 
from individual calorimeter cell energies in the calorimeter.
It is corrected for the presence of any muons and 
all energy corrections to leptons or to the jets  are propagated to \MET . 
Both experiments identify events with instrumental \MET\ by comparing missing
transverse energy calculations based on either reconstructed tracks or
calorimeter deposits.  The CDF
experiment employs an algorithm that combines tracking and calorimeter
information to improve \MET resolution.

\section{The LHC, ATLAS, and CMS experiments}

The LHC accelerator is a proton-proton collider operating at the highest
energies currently attained by a hadron collider, which started operation in 2010.  
During the 2011 data taking period
it was configured to collide beams with 7 TeV center of mass energy and provided
an integrated dataset of approximately 5~fb$^{\rm -1}$ to the LHC experiments, while during 2012 it was
configured at 8 TeV and provided an integrated dataset of approximately 5.8 (5.3)~fb$^{\rm -1}$ to ATLAS(CMS).  
The instantaneous luminosity of the LHC has reached 7.7x10$^{\rm 33}$ cm$^{\rm 2}$ s$^{\rm -1}$ making
20-30 multiple interactions per crossing a typical occurrence and
creating additional challenges for triggering on and
reconstructing physics events.  A proton-proton 
collider at high energy provides large cross sections for gluon-gluon or quark-quark
initiated Higgs boson production processes such as gluon fusion and
vector boson fusion.  For instance, the cross section for gluon fusion to a Higgs boson
is increased by a factor of approximately 15 compared to the Tevatron.
In addition, the increase in center of mass energy from 7 to 8 TeV
correspondingly raises the Higgs boson cross section by an additional factor of approximately 30\%.

The LHC experiments are forward-backward and cylindrically symmetric detectors
with tracking, calorimetric and muon detector elements.

The ATLAS detector includes an inner tracking and vertexing system,
electromagnetic and hadronic calorimetry and an outer
muon detection system~\cite{atlasexp}.  The inner tracking detector consists of a silicon pixel detector, 
a silicon micro-strip detector, and a transition radiation tracker immersed in the field of
a 2 T solenoidal magnet which provides charged particle tracking and vertex finding
over a large pseudorapidity range of $|\eta|<2.5$.  
The inner tracker and solenoid is surrounded by a high-granularity 
liquid-argon sampling electromagnetic calorimeter which provides electron (photon)
finding in the range $|\eta|<2.47$ ($|\eta|<2.37$).  An iron-scintillator tile calorimeter provides 
hadronic coverage in the central rapidity range. The end-cap and forward regions are instrumented 
with liquid-argon calorimetry for both electromagnetic and hadronic measurements which extends jet
finding to $|\eta|<4.9$. The muon spectrometer surrounds 
the calorimeters and consists of three large superconducting toroids, each with eight coils, 
a system of precision tracking chambers, and detectors for triggering in the range  $|\eta|<2.4$.

The CMS detector consists of a barrel assembly and two endcaps, comprising, in 
successive layers outwards from the collision region, a silicon pixel and 
strip tracker, a lead tungstate crystal electromagnetic calorimeter, a 
brass-scintillator hadron calorimeter, a superconducting solenoid,  
and gas-ionization chambers embedded in the steel return yoke for the detection of muons~\cite{CMS:2008zzk}.
The silicon detector provides 
charged particle tracking
and vertexing for $b$ tagging over a large pseudorapidity range of
$|\eta|<2.5$, which is well matched to the coverage of the barrel and
electromagnetic calorimeter as well as that of the muon chambers
providing coverage to $|\eta| < 3$ and $|\eta|<2.4$ for electron and muon
identification respectively.
The return field of the magnet allows independent momentum
measurement and triggering in the muon chambers.
The identification of photons and $\tau$ leptons in 
hadronic decay modes is performed within the overlapping pseudorapidity range
of the tracker and electromagnetic calorimeter.  
Jet finding can be performed in an expanded
pseudorapidity range up to $|\eta|<5.0$ using forward calorimeters.

Further detail about the primary systems of the experiments are given next.

\subsection{Tracking detectors}
The ATLAS tracker consists of a silicon pixel and strip tracker and a transition radiation straw tube
tracker immersed in a 2.0 T magnetic field provided by a solenoidal magnet.  The strip tracker consists of 
four barrel layers and nine end-cap disks at each end.  The pixel detector consists of three barrel layers and two end-cap
disks at each end.   The silicon inner detector tracks charged particles over the range  $|\eta| <2.5$ 
typically adding three pixels hits per track and allowing for  high efficiency 
primary vertex finding and $b$-jet identification.   The transition radiation tracker consists of 4 mm drift
tubes configured in a barrel region and two sets of multiple wheel
endcaps.  This configuration provides high efficiency $r-\phi$ tracking with typically 36
hits per track in the range $|\eta| <2.0$.  Between the straws a gas
mixture primarily consisting of xenon and carbon dioxide causes
ultrarelativistic charged particles to produce transition radiation photons
which are converted to electron-positron pairs in thin foils and leave much larger signals in some straws. 

The central feature of the CMS detector is a 6 m internal diameter 3.8 T solenoidal magnet.  The solenoid provides
a uniform bending field for a large volume silicon strip and pixel tracker. The strip tracker consists of 
ten barrel layers and six end-cap disks at each end.  The pixel detector consists of three barrel layers and three end-cap
disks at each end.  The  tracker typically allows 12 or more hits to
be found for all tracks, including three layers
of pixel hits, and allows for high efficiency tracking for $|\eta| <2.5$ as well as high efficiency 
primary vertex finding and $b$-jet identification.

\subsection{Calorimeters}
The ATLAS calorimeter uses sampling technologies over the entire angular area of coverage.
The electromagnetic calorimeter is a lead-liquid argon (LAr) sampling calorimeter
and is divided into overlapping barrel and end-cap portions with coverage up to $|\eta|<3.2$.  The
calorimeter is divided into three barrels and two end-cap parts, with very fine granularity
in their inner layers providing excellent angular resolution for electrons and photons.  
The barrel calorimeter has an extra inner LAr layer that functions as
a preshower detector with high longitudinal segmentation.
The mass resolution for diphotons with an invariant mass of 125 GeV
is approximately 1.5 GeV.   The system has
sufficient angular pointing ability to loosely associate photons with a primary interaction vertex to
reduce combinatoric backgrounds.
The hadronic sampling calorimeter is divided into a barrel region and endcaps that cover the range up to $|\eta|<3.2$
with seven to ten interaction lengths of instrumented material.  The barrel region is composed of steel absorber plates with
scintillating tile readout.  The end-cap calorimeters are based on LAr technology with copper absorber.  A forward
calorimeter based on copper (inner) and tungsten (outer) sampling extends the pseudorapidity range to $|\eta|< 5.0$ .  
The inner copper portion gives the forward calorimeter the ability to perform electromagnetic calorimetry.

CMS incorporates a lead tungstate crystal electromagnetic calorimeter and a brass and scintillator
sampling hadronic calorimeter enclosed within the solenoid.  
The electromagnetic calorimeter is divided into a barrel region and two endcaps
which extend coverage up to $|\eta|<3.0$.  The crystals have cross-sectional areas of 22x22 mm$^2$  in the barrel
and 29x29 mm$^2$ in the endcap, and the calorimeter material has a Moliere radius of 21 mm leading to narrow
showers and good angular resolution.  The end-cap calorimeter has two layers of silicon detectors interleaved
with lead layers configured as a preshower detector.  The excellent
energy and good angular resolution of the calorimeter
gives a diphoton mass resolution of 1.1 GeV at a mass of 120 GeV.
The hadronic sampling calorimeter is divided into a barrel region and endcaps that cover the range up to $|\eta|<3.0$
with 7 to 11 interaction lengths of instrumented material.  In
addition there is a tail catcher calorimeter located outside the solenoid which increases the material of the
calorimeter to at least ten interaction lengths everywhere.  An iron forward calorimeter readout
by quartz fibers extends the pseudorapidity range to $|\eta|< 5.0$ to improve measurement of transverse energy
and detection of missing transverse energy.

\subsection{Muon systems}

The ATLAS muon system  is based on the magnetic deflection of muon tracks within large superconducting 
air-core toroid magnets, instrumented with separate trigger and high-precision tracking chambers. 
The magnetic field is generated by three toroids, one in the barrel region,
$|\eta|<1.4$, and two in the end-cap regions, $1.6 < \eta < 2.7$, with an overlapping
region in between.  This magnet configuration provides a field which is mostly orthogonal to the muon trajectories
over the entire $\eta$ range.  Precision tracking is provided by monitored drift tubes
and cathode strip chambers, while resistive place chambers and thin gap chambers provide fast
triggering capability based on independent measurements of the particle momentum for the central and
forward regions respectively.  

The CMS muon system makes use of three technologies interleaved in the steel return yoke of the magnet; 
drift tubes (central) and cathode strip chambers (forward)
for precision tracking and triggering based on tracking information and resistive
plate chambers for tracking and triggering based on time measurements.
The return field of the solenoid saturates the return yoke providing a bending field for independently
measuring muon momenta.  The chambers are divided into four stations with 14 layers of
drift tubes or six layers of cathode strip chambers for robust tracking.

\subsection{Triggering}
Both detectors use a multilevel triggering system.
The first level of the trigger allows for the measurement of the momentum or energy of physics
objects including electrons and photons as electromagnetic energy deposits, muons with independent measurement
of the momentum in the muon systems, jets and missing transverse energy using full calorimeter information,
and taus as narrow jets.  The event rate is reduced to approximately
100 kHz at the level one. 
The ATLAS experiment employs a second level trigger which repeats physics object identification using
the full granularity of the detector and further reduces the event rate to 3.5kHz.  Both detectors
employ a full event reconstruction using optimized versions of offline reconstruction code
running on commercial processors as a level three or high level event filter with an output
rate of about 200 Hz  for ATLAS and 500 Hz for CMS.

\subsection{Physics object identification at the LHC}

Both experiments classify observed signatures in their detectors as physics objects
that can be associated with the particles and decay products of
particles produced in high-energy collisions.  Physics
analysis can be performed directly on these physics objects.  The various physics signatures
identified by the experiments are discussed next.

\subsubsection{Charged lepton identification}

For electron identification both experiments use clusters formed
in the electromagnetic calorimeters and associate them  to tracks found in the tracker 
by matching to their extrapolated position and energy~\cite{cms_electron1,cms_electron2,Aad:2010yt}.   
The clustering algorithm takes into account the typical spread
of the cluster in $\varphi$ due to bremsstrahlung photons.
Tracks are generally identified using inside-out
algorithms since the inner pixel detectors are the least occupied tracking system due to their high
granularity.  Electron candidates are required to pass requirements on cluster shape information, energy
leakage information in the hadronic calorimeter, and track quality information.
Different operating points are defined for different levels of
selection efficiency and background rejection.  
Tighter operating points with lower efficiency and better background rejection include tighter
criteria on identification requirements, requirements to reject electrons from photon
conversions, and requirements on track impact parameter to reject electrons from interactions with
matter or long=lived decays.
The CMS experiment employs a BDT based multivariate electron identification
algorithm for several analyses to improve efficiency and background rejection.

Because of the large bending fields in the muon spectrometers of the LHC experiments, muons can be
identified in the muon systems independently of the tracker~\cite{Aad:2010yt,cms_muon1,cms_muon2}.
In addition muons can be identified in
the combined muon and tracking systems.  Both experiments form combined tracker and muon system muons
by associating independently reconstructed muons in the muon systems to charged tracks in the tracker
using position and momentum information.  CMS additionally identifies muons by extrapolating tracks
into the muon system to perform an inside-out search for compatible muon system hits, which improves muon
finding efficiency
at low transverse momenta.  Muons used in analysis are generally required to be identified in both
systems, pass minimum tracker hit and muon system segment
requirements, and satisfy requirements on track impact parameter
to help reject decays to muons from long lived particles and cosmic rays.

Both experiments increase the purity of identified electrons and muons by requiring that the charged lepton candidates
be isolated, which rejects real and misreconstructed charge leptons contained within jets.  The ATLAS experiment applies relative
calorimeter based isolation for electrons and relative tracker based isolation for muons based on
the total calorimeter energy or track momenta  found in a cone around the candidate divided by the transverse energy or momentum of the
candidate.  The CMS experiment applies relative particle flow based isolation based on charge tracks;
electromagnetic energy from electrons, photons or neutral pions; and neutral hadronic energy not associated
with tracks found in a cone around the candidate and divided by the transverse energy or momentum of the candidate.
These ratios are required to be less than a given value which can be adjusted to achieve different
levels of performance.  The particle flow technique as used at CMS is described in more detail
at the end of this section.

The identification of hadronically decaying $\tau$ leptons is characterized by the presence of one or
three charged hadrons, identified as tracks with associated calorimeter
energy, and possible narrow strips of electromagnetic energy
deposits characteristic of neutral pion decay to photons, all contained in a narrow collimated jet~\cite{cms_tau}.  
The ATLAS experiment combines this information together in a boosted decision tree based multivariate 
discriminant~\cite{atlas_tau}.  
The CMS experiment uses a particle flow technique to measure
the charged hadrons in the tracking detector and neutral pions as
strip shaped electromagnetic energy deposits.  It also improves the mass resolution of objects reconstructed 
as a hadronically decaying $\tau$ by performing a fit constraining the objects from the $\tau$ decay to the $\tau$
lepton mass.

\subsubsection{Photon identification}

The identification of photons uses similar criteria to those used for electrons except that
events with a track or track segment compatible with the electromagnetic cluster are rejected.  
Photon candidates are formed
from electromagnetic clusters in the EM calorimeter~\cite{atlas_photon}.  
The clustering algorithm and subsequent identification
criteria allows for the possibility that the photon converts to an electron pair.  Photon candidates are required 
to pass requirements on cluster shape information and energy
leakage information in the hadronic calorimeter.  The ATLAS experiment additionally applies isolation requirements,
while the CMS experiment includes the above criteria and isolation information in a BDT
algorithm designed to reject non prompt sources of photons.  The ATLAS experiment also
uses the longitudinal segmentation of its EM calorimeter to require that the photons are compatible with
pointing to  the primary high transverse momentum interaction vertex.  In cases of conversion early in the material
of the tracker, both experiments reconstruct the electron-position conversion pairs when possible.  In this case
the vector sum of the track momenta are required to point toward the primary interaction vertex.

\subsubsection{Light and heavy flavor jets}
Jets are generally reconstructed using the anti-$k_t$ algorithm based on calorimeter clusters~\cite{Cacciari:2008gp}.
The excellent granularity	
of the LHC detectors allows for the effective use of such an iterative clustering-based jet-finding algorithm.
Raw jet energy measurements are corrected for imperfect calorimeter
response using correction factors from studies of detector response in
test beam data, MC simulations, and collision data.  
In high pileup conditions jets can be required to point
to the hard interaction vertex.
Heavy flavor jets, or jets originating from $b$ quarks,  
are found based on the positive track impact parameter significance of tracks and reconstructed secondary
vertices~\cite{cms_bjet,atlas_bjet}.  
The ATLAS experiment combines the track impact parameter significance information to form a likelihood
ratio quantifying whether a jet originates from a $b$-flavored parton
or light parton.  A second likelihood is formed using secondary vertex information including decay length
significance and vertex mass information.  
The further use of a likelihood or similar technique allows the results of the two types of algorithms to be combined into a
single continuous $b$-flavored jet identification variable.
The experiments define multiple operating points with different
selection efficiencies and background rejection.
As an example using this type of information both experiments can achieve 
50\% efficiency with a factor of 1000 in background rejection.

\subsubsection{Missing transverse energy}
Both experiments use measurements of missing transverse energy  
to identify events with neutrinos~\cite{cms_jetmet,Aad:2012re}.
The ATLAS experiment measures the visible energy using electromagnetic clusters corrected 
for the average hadronic component
and corrected for the transverse momentum of identified muons.
The magnitude and direction of the missing transverse
energy are obtained from the energy imbalance in the transverse plane.
The CMS experiment uses a a particle
flow method to measure the visible energy and infers the \MET\
measurement in the same way.  The \MET\ measurement can
be improved in the presence of high pile-up by using associated objects to calculate the \MET\ for a given vertex.  The
excellent tracking efficiency and z coordinate resolution of the experiments are essential for this technique.  
In addition events with false \MET\ from jet mismeasurement are identified by methods such as checking 
whether the \MET\ is collinear with
jets or charged leptons or comparing different methods of the \MET\
measurement such as those based solely on reconstructed charged tracks or
calorimeter energy cluster information.

\subsubsection{Particle flow}

In CMS, jet and missing transverse energy reconstruction, and $\tau$ lepton identification 
are substantially improved by using
particle flow identification techniques~\cite{cms_pf1,cms_pf2} 
that classify detector signatures as charged or neutral hadrons,
photons, or charged leptons using combined information from the tracker, calorimeters, and muon
detectors.  Electrons, photons, and neutral pions are measured in the EM calorimeter.  Muons are measured
in the tracker and muon detectors. Tau leptons in decay modes involving hadrons are found combining tracker
information for charged hadrons and EM information for neutral pions.  Charged hadrons in jets are measured
using the tracker. Neutral energy not associated with any of the above objects is measured as energy clusters
in the EM and hadronic calorimeters.  The vector sum of particle flow objects can also be used to identify
missing transverse energy.

\section{Tevatron low-mass Higgs boson searches}
At lower masses ($m_H < 135$ GeV), the dominant decay of the Higgs boson is 
$H \rightarrow b\bar{b}$, but it is hopeless to search for direct 
($gg \rightarrow H$) Higgs boson production
in this decay mode, due to the overwhelming multijet background. 
However,  $q \bar{q}$ annihilation results in
associated vector boson-Higgs boson production ($VH$) in $p \bar{p}$ collisions with a better signal-to-background
ratio than available at a $pp$ collider.
The ``primary'' channels for searching for a low-mass Higgs boson at the Tevatron,
$WH$ and $ZH$ production, are best studied in the $\ell\nu b\bar{b}$, 
$\ell \ell b\bar{b}$ or $\nu \nu b\bar{b}$ final states. 
The lower branching ratios or poor signal to background of the other decay modes
render their sensitivity smaller than these primary channels but are
also searched for to provide additional sensitivity in the combination of all channels.

The associated production analyses generally proceed 
in three steps. The first is preselection where a high statistics
sample of events containing bosons and jets is constructed, allowing
for detailed validation of background modeling.
The $W$ and $Z$ are required to decay leptonically [topologies where $W$ or $Z$ decay
hadronically are also searched for, but are less sensitive~(\cite{cdfjjbb})]
to facilitate event
triggering and selection. Electrons and muons (including those coming from taus decaying
leptonically) allow for a relatively pure selection, but $Z \rightarrow \nu\nu$ decays are
also exploited.
In the next selection step, at least one jet is required to be identified as 
a $b$-quark jet,
enhancing the signal to background in separate final subsamples
defined by the 
number and type of $b$-quark jets found.
In the final step, multivariate analysis 
techniques are applied to each
of the samples to further separate the potential signal from the backgrounds.

Vector boson fusion production, associated production with vector
bosons,  and direct Higgs boson production
can also be exploited when the Higgs boson decays to a $\tau\tau$
pair, by making use of the kinematics of the
potential additional jets in the final state. These processes suffer from significant background and
so are considered as secondary channels at the Tevatron. Another secondary channel which is exploited
at the Tevatron is inclusive production of a Higgs boson with
Higgs boson decay to two photons. The sensitivity
of this channel is low due to the small branching ratio of this decay, typically smaller than 0.2~\%.
The $t \bar{t} H$ production is also searched for, but has low sensitivity.

In all the analyses, the data are separated into multiple orthogonal search samples of varying 
sensitivities. The analyses are described next and all results are summarized in section
IX.

	    \subsection{{$VZ$} with { $Z \rightarrow b\bar{b}$} as a test of the 
{$VH$} search}
	
Since the low-mass analyses use advanced multivariate techniques for
separating signal and backgrounds and are obtained from low
signal-to-background search samples, a crucial test has been performed
considering $WZ$ and $ZZ$ diboson production with $Z$ decays to heavy flavor
as signal,  
mimicking the final states of $WH$ and $ZH$ including the essential
feature of  resonant dijet production. In such analyses
the $WW$ process is taken as a background, with a normalization constrained to its NLO
cross section. These analyses, and their combination, are performed in the same way as their 
Higgs counterparts. The CDF+D0 combination displays strong evidence (4.6 $\sigma$)
for such production, with a measured cross section $\sigma(VZ)=4.47\pm 0.97$ pb consistent
with the SM prediction~\cite{comb-diboson}. Evidence was first seen by both collaborations
separately~\cite{cdf-diboson,d0-diboson} using 9.5 and 7.5 fb$^{\rm -1}$ respectively.

	\subsection{{$WH \rightarrow l \nu b \bar{b}$}}

The search for the process $q\bar{q}\rightarrow WH + X$ in which a quark-antiquark pair 
leads to the production  of the Higgs boson in association with a $W$ boson
is based on a total integrated luminosity $\cal{L}$ $\simeq 10 ~\rm fb^{\rm -1}$ of 
collision data collected by both the CDF and D0 detectors at the Fermilab Tevatron  $p\bar{p}$ collider 
between 2002 and 2011 ~\cite{CDFWH3,D0WH3,CDFWHprel,D0WHprel}. 
Candidate $W$ boson events are preselected via their 
decays to an electron or a muon plus a neutrino ($W\rightarrow e$ or $\mu \nu$) 
while the Higgs boson is identified through its decay mode into a pair 
of $b$ quarks ($H\rightarrow b\bar{b}$).
The experimental signature is a single 
isolated lepton, missing transverse energy, and either two or three 
(to accommodate additional gluon radiation in the 
hard collision) jets, at 
least one of which is required to be consistent with having been
initiated by a  $b$ quark.

\begin{figure}[t]
\vskip -0.4cm
\hspace{-1.0cm} \includegraphics[width=2.7in,height=2.0in]{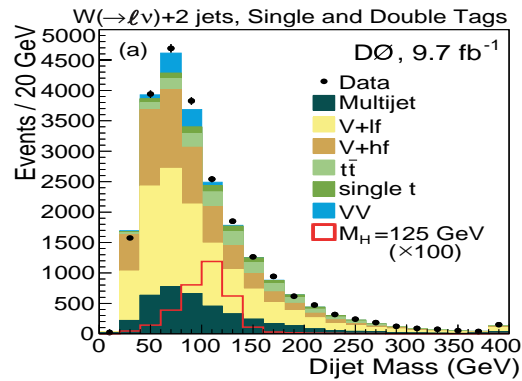}
\caption{Dijet mass distribution in the D0 $WH$ analysis for the $W+2$ jet sample with 1 or
2 identified $b$-jets.
The data are well described by the sum of all the SM backgrounds. The simulated
signal is also represented.}
\label{dijet}
\end{figure}

To increase signal acceptance, the lepton identification criteria are
as loose as possible.
This results in backgrounds originating
from MJ events, in which one of the jets is misidentified as an isolated
lepton.
In the CDF analysis, the  MJ background 
is strongly reduced by kinematic cuts and by using a dedicated multivariate technique
to reject this background~\cite{CDFWH3}. The remaining
MJ background  contribution is modeled from the data using side-band techniques. 
In the D0 analysis, the MJ background contributions passing the preselection 
criteria in each sample are determined from the data using an
unbinned matrix method approach~\cite{D0WH3}. 
The ``physics'' backgrounds with similar event topologies are modeled using
Monte Carlo event generators. 
The SM predictions are used to set the relative normalizations of 
all of the generated samples, with additional normalization factors applied to samples 
of $W$ bosons $+$ $n$ partons generated using the {\tt ALPGEN} Monte Carlo event generator. These factors are
determined at the preselection stage where the SM Higgs boson
contribution is negligible. The predicted backgrounds model the
data well in the high statistics sample before a $b$-tagging algorithm is applied, as shown in~Fig.~\ref{dijet}.

The CDF and D0 analyses proceed by subdividing the selected sample into orthogonal 
subsamples based on how many of the jets in the event, one or two,  are consistent with 
having been initiated by a heavy $b$-quark, and at what level (``loose ($L$)'' or ``tight ($T$)'')
of confidence. CDF has five tagging categories
($TT$,$TL$,$LL$,$T$,$L$) for the two jet sample, and
two categories ($TT$,$TL$) for the three jet sample, while D0 uses four categories
($TT$,$TL$,$LL$,$T$) and two categories ($LL$,$T$) respectively.
In two $b$-tagged jet events, the
dominant remaining backgrounds are from $Wb\bar{b}$, $t\bar{t}$, and
single top-quark production.
In single $b$-tagged jet events the dominant backgrounds 
are $W +$  light or $c$-quark jet production as well as MJ 
background events.
To further discriminate the  remaining backgrounds from the signal, MVA techniques are applied to each  subsample.
Some of the discriminant variables used in these analyses 
are given in Table~\ref{rf_variables}. 

\begin{table}
\begin{tabular}{|l|l|}
\hline 
rf input variable & Description\\
\hline
\MET{} 				 & Missing transverse energy \\
$M_W^T$ 			 & Lepton-\MET{} transverse mass  \\
$p_{T}$($\ell$-\MET{} system)    & $p_{T}$ of $W$ candidate\\ 
$p_{T}$($j_1$) $p_{T}$($j_2$)  & Leading (subleading) jet $p_{T}$ \\
$m_{jj}$ 			 & Dijet invariant mass \\
$p_{T}$(dijet system) 	 	 & $p_{T}$ of dijet system \\
$\Delta\mathcal{R}$($j_1$,$j_2$) & $\Delta\mathcal{R}$ between jets \\
$\Delta \phi$($j_1$,$j_2$) 	 & $\Delta \phi$ between jets \\
$H_{T}$		            	 & Scalar sum of $p_T$ of all jets\\
\hline
\end{tabular}
\caption{\label{rf_variables} Description of some characteristic kinematic input 
quantities of the MVA technique used in the $WH$ analyses.}
\end{table}

Systematic uncertainties affect not only the normalization of the signal and backgrounds, 
but also the shape of the MVA
output distributions. The influence of each source of systematic uncertainty
is studied separately for each of the independent subsamples.
Uncertainties in the efficiencies of selection, on jet calibration, and on the $b$-tagging criteria 
affect the precision at which the background modeling is known.
The uncertainties on the parton density functions and  
the effect of renormalization and factorization scales on signal and background  simulation are also 
taken into account. All these uncertainties are allowed to affect the shape of the MVA output distributions.

	\subsection{$ZH \rightarrow llb\bar{b}$}

The search by the CDF and D0 collaborations at the Tevatron
for the process $q\bar{q}\rightarrow ZH + X$ in which a quark-antiquark pair 
leads to the production  of the Higgs boson in association with a $Z$ boson decaying
to a pair of charged leptons
is also based on a total integrated luminosity $\cal{L}$ $\simeq 10 ~\rm fb^{\rm -1}$~\cite{CDFZH3,D0ZH3,CDFZHprel,D0ZHprel}. 
Candidate $Z$ boson events are preselected via their decays into $e^+e^-$ or
$\mu^+\mu^-$ pairs, and the associated Higgs boson is identified through its decay into a pair 
of heavy $b$-quarks ($H\rightarrow b\bar{b}$).
Candidate events are required to have two or three jets, at least one of
which is identified as a $b$ jet.

In this final state, which requires two leptons, the MJ background is
negligible.
The physics backgrounds are modeled using the 
same Monte Carlo event generators used in the $WH$ analysis.

To maximize the lepton acceptance and benefit from higher quality lepton categories, the events are classified
according to the lepton types. Those having both leptons identified with high confidence 
are treated separately from the others which contain 
loosely identified, forward, or track-based leptons. These samples are analyzed independently, 
allowing for an optimal sensitivity of the search.
In addition, multivariate lepton selections are used.
In CDF, to enhance the discriminating power of the dijet
invariant mass, 
a NN derived energy correction is applied to the jets. 
This correction depends on the missing 
transverse energy and its orientation with respect to the jets. In D0, jet energy
resolution improvements are obtained through a kinematic fit of the complete event, 
since all particles can be detected in this process.

These analyses 
also proceed by subdividing the selected sample into orthogonal 
subsamples based on the number and the quality of the $b$-tagged jets
in the event.
CDF has four tagging categories ($TT$,$TL$,$LL$,$T$) for both the $Z+2$ and $Z+3$ 
jet samples,
while D0 has a different treatment using two tagging categories ($TL$,$T$) for the $Z+3$ jet sample.
CDF
employs two NNs to simultaneously separate signal events from the dominant 
$Z+$jets and kinematically different $t\bar{t}$ backgrounds. These NNs use various kinematic distributions, 
matrix element probabilities, and the output of a separate jet flavor separating NN as inputs.
In single and double $b$-tagged jet events, the
dominant remaining background is $Zb\bar{b}$.
To suppress the remaining background MVA techniques are applied to each  subsample.
Systematic uncertainties are overall less important than for $WH$ since no missing transverse energy is involved,
but most of the other systematic uncertainties are of similar
magnitude to those of the $WH$ analyses.

	\subsection{$ZH \rightarrow \nu\nu b \bar{b}$ and $VH \rightarrow \MET b\bar{b}$}

The remaining $b\bar{b}$  analysis is built to detect the $ZH \rightarrow
\nu\nu b \bar{b}$ process but is also sensitive to  $WH$ events in which the
charged lepton is not identified, hence its alternate label as 	
$VH \rightarrow \MET b\bar{b}$.
These searches are based on a total integrated luminosity 
$\cal{L}$ $\simeq 10 ~\rm fb^{\rm -1}$~\cite{CDFvvbb3,D0vvbb3,CDFvvbbprel,D0vvbbprel}.
Since this final state contains no leptons, triggering on these
events and modeling the effects of the trigger requirements on the event selection
are significant challenges.
Both CDF and D0 use triggers based on \MET , with or without accompanying jets.
The analyses are performed while studying in parallel several control samples to monitor
the understanding of the background. 
Events are required to have significant  \MET\ and two or three jets, well separated
from the \MET\ direction. For the preselection, multivariate approaches are also applied to the events
to remove a large part of the MJ background. For the final selection
$b$ tagging is employed.

For this analysis, the preselection plays a crucial role, given the size of the MJ background.
As an example, at D0, the preselection uses the following main requirements.
The events must have a well-reconstructed interaction vertex  and
two or three jets with associated tracks to ensure efficient operation of the
$b$-tagging algorithm.
These jets must have $ p_T >$ 20 GeV and $|\eta| <$ 2.5 
and not be back-to-back in the transverse plane. \MET\ must be greater than 40 GeV,
with large significance, i.e. with \MET\ values that are less likely
to have originated from fluctuations in jet energies,
and the scalar sum of the two leading jet $p_T$ must be greater than 80 GeV.

The dominant signal topology is a pair of $b$ jets recoiling against the \MET\ due to the neutrinos from $Z$ decay
with the direction of the \MET\ at large angles relative to both jet
directions.  Conversely, in the case of events from MJ background with
fluctuations in jet energy measurements, the \MET\ tends to be aligned with a mismeasured jet. An alternate estimate
of \MET\ can be obtained from the missing $p_T$ calculated from the reconstructed charged
particle tracks. This variable is less sensitive to jet energy measurement fluctuations and, in signal events, is also
expected to point away from both jets, while in the MJ background its angular distribution is expected to be more
isotropic. A variable characterizing these features is used to further reject the MJ background.
As for the other $b \bar{b}$ analyses, the physics backgrounds are taken into account using
Monte Carlo event generators. 

\begin{figure}[t]
\vskip -0.4cm
\hspace{-1.0cm} \includegraphics[width=3.2in]{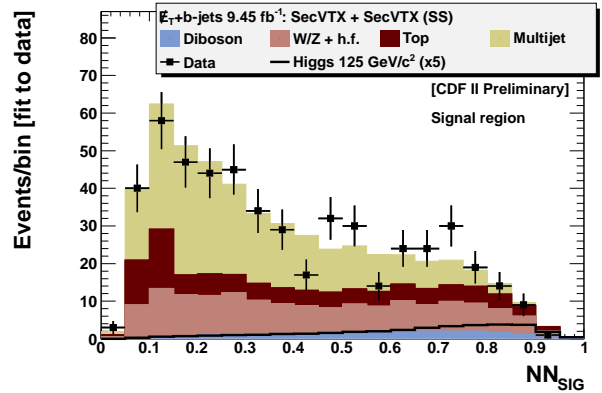}
\caption{ 
Distribution of the NN discriminant in the CDF $\MET bb$ analysis for
the sample with two or three jets where 2 of the jets are tagged as
$b$ jets
by the secondary vertex algorithm.
The data are well described by the sum of all the SM backgrounds. The simulated
signal is also represented.}
\label{metbbnn}
\end{figure}

The preselected samples are subdivided into orthogonal 
subsamples based on the number and the quality of the $b$-tagged jets in the
event.  CDF has three tagging categories ($SS$,$SJ$,$S$) for the analyzed two and three
jet samples,
where $S$ represents a jet identified by a reconstructed secondary vertex, and
J represents a jet identified by the presence of tracks not pointing to the
main interaction vertex. D0 has two tagging categories for its two jet
sample ($TT$ and $LL$ or $T$).
To suppress the remaining backgrounds multivariate
discriminant techniques are applied to each subsample as in Fig.~\ref{metbbnn}.


\subsection{CDF and D0 results on $H\rightarrow b \bar{b}$ searches}

Here we present the individual results from each collaboration.
The results are extracted using the MVA
discriminant distributions from each subchannel of  these three $H\rightarrow b \bar{b}$ analyses, 
and then combined. The CDF+D0 combination is discussed in Sec. IX. Here we
present the results of each collaboration, while
the limits for the three different search topologies are given in Table~\ref{tab:cdfacc}.
The statistical techniques used are described in Sec. IX and allow for the extraction of 
the limit on the signal cross section normalized to the SM expectation, 
or, in case of excess, determine the $p$-value of the background
fluctuation.
At very low mass, 
when combining the three $H\rightarrow b\bar{b}$ topologies,
CDF and D0 
exclude at 95\% C.L. Higgs bosons with masses smaller than 96 and 102~GeV, respectively.
However, in the results from both collaborations an observed limit
above the background-only expectation is obtained for the $\simeq$ 120-140 GeV
range. In particular, the expected or observed limits at $m_H = 125$ GeV are
1.8 or 4.2 and 2.3 or 3.2 times the SM expectation,  for the CDF~\cite{cdf-bb-comb} 
and D0~\cite{d0-bb-comb} searches, respectively. To quantify the excess,   
the local $p$ values are calculated and found to be minimal 
for a Higgs boson
mass of 135 GeV at CDF and D0, where the local significance of these
deviations with respect to the background-only hypothesis
is 2.7$\sigma$ (1.7$\sigma$), which themselves correspond to  2.5 $\sigma$ (1.5$\sigma$) global significances
after applying look-elsewhere factors (cf. Sec. IX). These two mass values are compatible
given the resolution of the dijet mass in these final states.


  \subsection{Searches in $\tau_{{h}}$ final states}

Higgs boson searches using tau leptons decaying hadronically  ($\tau_h$)
complement those using electrons and muons.
CDF performs a generic analysis searching for Higgs bosons decaying
to $\tau$ lepton pairs originating from direct $gg \rightarrow H$
production, associated $WH$ or $ZH$ production, and vector boson fusion production~\cite{cdfHtt}.
A final state consisting of one leptonic $\tau$ decay and one hadronic 
$\tau$ decay or two leptonic $\tau$ decays of different flavors $e\mu$ is required.
CDF hadronic $\tau$ identification is performed using an MVA approach.
The final discriminant for setting limits is obtained combining the output of
four MVAs  trained to separate a potential signal
from each of the four primary backgrounds ($Z \rightarrow \tau \tau$, 
$t \bar{t}$, multijet, and $W$+jet production).  
CDF also has an analysis of events that contain one or more reconstructed 
leptons ($\ell$ = $e$ or $\mu$) in addition to a $\tau$ lepton pair focusing on 
associated production where $H \rightarrow \tau \tau$ and additional leptons 
are produced in the decay of the $W$ or $Z$ boson~\cite{cdfVHtt}. Events 
are separated into five 
separate analysis channels ($\ell \ell \ell$, $e \mu \tau_{{h}}$, $\ell 
\ell \tau_{{h}}$, $\ell \tau_{{h}} \tau_{{h}}$, and $\ell \ell 
\ell \ell$). The four lepton category includes $\tau_{{h}}$ candidates.  The 
final discriminants are likelihoods based on outputs obtained from independent 
MVA trained against each of the primary backgrounds ($Z$+jets, $t\bar{t}$, and 
dibosons).  

The D0 $\ell \tau_{{h}}jj$ analyses also include direct $gg 
\rightarrow H$ production, associated $WH$ or $ZH$ production, and vector boson fusion 
production~\cite{dzVHt2,dzttl}.  Decays of the Higgs boson to tau, $W$, and $Z$ boson pairs are 
considered. A final state consisting of one leptonic $\tau$ decay, one hadronic 
tau decay, and two jets is required. Both muonic and electronic subchannels 
are considered. 
The outputs of boosted decision trees are used as the final discriminant.

	    \subsection{Searches in $H \to \gamma \gamma$}

Both CDF~\cite{cdfHgg} and D0~\cite{dzHgg} searched for Higgs bosons decaying into
diphoton pairs with the full statistics (10 fb$^{-1}$).  
The CDF analysis searches for a signal peak in the diphoton
invariant mass spectrum above the smooth background originating from QCD production
in several detector based categories with different signal-to-background ratios.  
In the D0 analysis the contribution of jets misidentified as photons 
is reduced by combining information sensitive to differences in the energy 
deposition from real or false photons in the tracker and in the calorimeter 
in a neural network output (NN$_o$). The output of an MVA, rather than the 
diphoton invariant mass, is used as the final discriminating variable. 
The final MVA takes as input variables the NN$_o$, 
the transverse energies of the leading two photons and the azimuthal 
opening angle between them, the diphoton invariant mass and transverse momentum,
and additional variables, bringing an improvement in 
sensitivity of $\approx 20\%$.

	    \subsection{Searches for $t\bar{t}H$ production}

The $t\bar{t}H$ production is interesting for the direct $tH$ coupling it involves, however
its cross section is too small at the Tevatron to contribute strongly
to the overall search sensitivity. 
CDF uses several nonoverlapping sets of events to search for 
the process $t \bar{t} H \rightarrow t \bar{t} b \bar{b}$.
Events with a reconstructed lepton, large missing transverse energy, and
four, five, and six or more jets are further subdivided into five $b$-tagging 
categories ~\cite{cdfttHLep}.
Neural network discriminants are used to set limits.  
Events with no reconstructed lepton~\cite{cdfttHnoLep} are  separated 
into two categories, one containing events with large missing transverse energy 
and five to nine reconstructed jets and another containing events with low missing 
transverse energy and seven to ten reconstructed jets.  
A minimum of two $b$-tagged jets is also required
and events with three or more $b$ tags are analyzed 
separately from those with exactly two tags.  Neural network 
discriminants are used to reject large MJ background contributions and
separate potential $t\bar{t}H$ signal events from $t\bar{t}$ background 
events.

\section{Tevatron high mass Higgs boson searches}

As the hypothesized source of electroweak symmetry breaking, the Higgs boson has
strong coupling to both massive electroweak bosons.  At Higgs boson
masses above 135 GeV the decay to a pair of $W$ bosons is dominant, but even
below the threshold to produce on-shell $W$ bosons, the decay rate
to one real and one virtual $W$ boson is substantial.  
Both experiments pursue
a strategy of searching for $H \rightarrow W^+W^{-(*)}$ decay in
final states with at least one charged lepton and from
all production processes with substantial cross section.  In addition, searches
for the subdominant decay $H \rightarrow ZZ$ are performed.   The high-mass
search has similar sensitivity to the individual searches for
associated Higgs boson production
with a $W$ or $Z$ boson performed using the $H \rightarrow bb$ decay
mode at a Higgs boson mass
of 125 GeV. As proof of principle the experiments have observed all of the direct diboson
production processes with pairs of heavy gauge boson in final states that
are topologically similar to those used in the Higgs boson search.

\subsection{Diboson analysis: $WW$, $WZ$ and $ZZ$}
At the Tevatron,  the diboson analyses in final states with charged leptons are
performed using the same techniques used in the high mass Higgs boson search.  
The diboson searches based 
on leptonic and semileptonic decay modes allow for testing analysis techniques and developing 
further understanding of the primary backgrounds to the Higgs boson search.  
The direct SM production of the $WW$~\cite{Abe:1996dw,Abazov:2004kc}, $WZ$~\cite{Abulencia:2007tu}, and $ZZ$~\cite{Abazov:2008gya} 
have been observed in leptonic decay modes with two charged leptons, three charged leptons and four charged leptons
respectively. 
The diboson searches have also been performed with larger datasets using the most modern lepton
selections, providing measurements of 
the $WW$~\cite{Abazov:2009ys,Aaltonen:2009aa} $WZ$~\cite{Aaltonen:2012vu,Abazov:2012cj} and 
$ZZ$~\cite{CDF:2011ab,Abazov:2012cj} cross sections. 
The combined production of $WW$ and $WZ$ boson pairs has been observed
in events with one charged lepton and jets~\cite{Aaltonen:2009vh} and such an approach 
has been applied to perform another high-mass Higgs boson search~\cite{Abazov:2008yg}.
The combined production of all pairings
of massive vector bosons has been observed in events with one vector boson
decaying leptonically and the other hadronically~\cite{Aaltonen:2009fd}.

\subsection{Analysis Topologies}

\begin{figure}
\includegraphics[width=0.5\textwidth]{{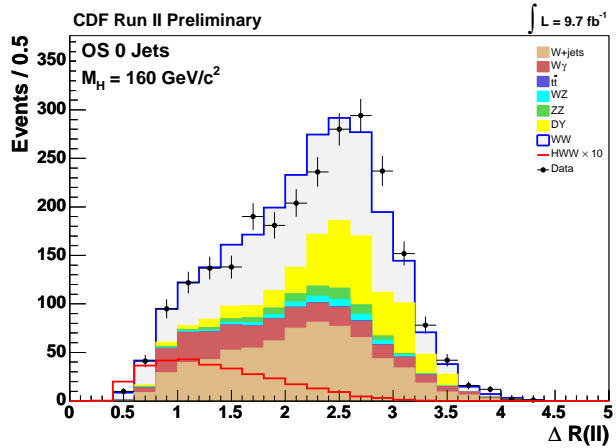}}
\caption{The angle between the lepton candidates in three dimensions:  
The distribution of this angle for background
and a $H\rightarrow WW$ signal with $m_h = 165$~GeV are compared}
\label{hwwdeltaR}
\end{figure}

  The analysis requirements typically require 
a triggered lepton with $p_T > 20$ GeV, possibly additional leptons with lower
thresholds,  and significant \MET\ that is not aligned along the direction of other physics objects in the events.
The events are then categorized into a large number of topologies that are
consistent with various Higgs boson production and decay modes.
These topologies are characterized by the number of charged leptons, whether the leptons
are the same or opposite charge, and the number of jets.   
Each topology involves a limited set of dominant signals and
backgrounds allowing for optimal discrimination, and
is therefore analyzed separately.  
The most sensitive analysis topology involving zero jets and leptonic
$H \rightarrow W^+W^-$ is described in detail below, while the
subdominant modes are briefly discussed afterward.
%
%
%

i)  {$ ggH \rightarrow W^+W^- \rightarrow \ell^+\nu\ell^-\bar{\nu} + n_{j}${jet}~\cite{cdfHWW,dzHWW}}. 
When $n_j$= 0, the signature is two opposite sign leptons, \MET, and no observed jets.
The signal in this final state is almost 100\% produced by the $ggH$ process.  
The dominant background is
from SM direct $WW$ production with minor contributions from Drell-Yan production, 
the $WZ$ and $ZZ$ diboson processes where one or more
charged leptons are not detected, and $W$+jets or $W$+$\gamma$ where a jet is misidentified as a lepton
or the $\gamma$ converts to an electron-positron pair, only one of which is detected.  The strongest discriminant
is the opening angle between the leptons in two dimensions (2D)
$\Delta\phi$ or three dimensions (3D) $\Delta R$ (Fig.~\ref{hwwdeltaR}), due to 
the spin correlation between the two spin one $W$ bosons  when decaying from the scalar Higgs boson.
The collinear topology of the charged leptons also results in a low dilepton invariant mass while Drell Yan
background peaks at the $Z$ mass and other backgrounds at large mass.
The neutrinos are also collinear leading to larger \MET.
Matrix element probabilities
are effective discriminants because in the zero jet topology the final state of either
Higgs boson or SM direct $WW$ production is well described by a leading order matrix element.  The
transverse mass of the Higgs boson can be well reconstructed since the
neutrinos are also collinear.
The D0 experiment further subdivides this mode by lepton flavor.
CDF subdivides this analysis into modes with two well-identified leptons, or
one well-identified lepton plus and an isolated track, and
analyzes events with low dilepton invariant
masses as a separate category.

When $n_j=1,2$, the signature  is two opposite sign leptons, \MET, and
observed jets.
These topologies have substantial contributions from $VH$ or VBF where
the jets are observed from either one of the vector boson decays or final state quarks
respectively and additional background from top pair production.

ii) {$VH \rightarrow V W^+W^- \rightarrow \ell^+ X \ell^+\nu+$X~\cite{dzWWW2,dzlll}.} 
Associated production events can result in events with either same
sign leptons or trileptons when the associated
vector boson decays to charged leptons.
The background includes $W$+jets with a misidentified lepton in the same
sign mode and SM direct $WZ$ production in the trilepton case.\\

iii) {$(ggH,VH,\mathrm{VBF})\rightarrow H\rightarrow WW \rightarrow \ell\nu$~$+\ge$ 2 jet.}
Higgs boson production can be searched for inclusively in events where one of the $W$ bosons decays leptonically and the
other $W$ boson decays to two quark jets~\cite{dzHWWjj}. 
The dominant backgrounds are from $W$+jets and multijet background where a jet
is misidentified as a lepton.  

iv) Other decay modes:
The CDF and D0 experiments also consider modes where one $W$ boson decays to a $\tau$ lepton which decays
hadronically~\cite{cdfHWW2}.  

Finally a search for the Higgs boson is performed in the $H \rightarrow ZZ$ mode where both $Z$ bosons
decay to charged leptons~\cite{cdfHZZ}.  The only significant background in this mode is SM direct $ZZ$ production.  The Higgs boson
can be detected by looking for a narrow resonance in the four lepton invariant mass distribution.  D0 includes
acceptance for $H \rightarrow ZZ$ in other cases where less than four charged leptons are found in the above
searches.

\begin{figure}
\includegraphics[width=0.45\textwidth]{{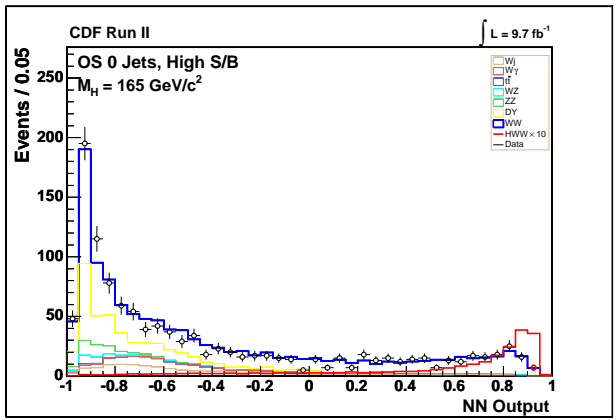}}
\caption{NN discriminant for   $ ggH \rightarrow WW \rightarrow \ell\nu\ell{\nu} + 0$ jet
at CDF. The distribution for background
and a $H\rightarrow WW$ signal with $m_H = 165$~GeV (multiplied by 10) are compared.}
\label{cdf_hwwnn}
\end{figure}

\subsection{CDF and D0 results at high Higgs boson mass}
All these search channels select a total of approximately 75 Higgs boson events 
per detector for a Higgs boson mass of
165~GeV.
More than half of these events are distinguished from the background
with a good signal-to-background ratio using MVA discriminants
as illustrated 
in Fig.~\ref{cdf_hwwnn}. 
Examining  the $\Delta R$ distribution (Fig.~\ref{hwwdeltaR}) shows that 
the largest discriminating power comes from the
spin correlation variable, due to the unique scalar nature of the Higgs boson.

No significant excess is seen in any of the high-mass Higgs boson search modes
and limits are thus extracted,
taking into account systematic uncertainties.
The theory uncertainties become larger in events with more jets since, for instance, 
in the NNLO calculation, events
with two additional jets are calculated only at NLO accuracy.  
These uncertainties are addressed following the treatment by~\cite{Anastasiou:2009bt,Campbell:2010gg}
and included in the limit extraction.

The expected limits with respect to the SM expectation for each experiment using the combination of all
high-mass Higgs boson search topologies
at $m_H = 165$~GeV are 0.69 and 0.72 times the SM Higgs boson cross section 
for the CDF~\cite{cdf_prel_comb} 
and D0~\cite{d0_prel_comb} searches, respectively.  
The experiments each achieve expected sensitivity within a factor of approximately 1.5 of the SM cross section, 
in the mass range $m_H = 140-185$~GeV.
The CDF (D0) analysis excludes the SM Higgs boson
in the mass range $m_H = 148-175$~GeV ($157-172$~GeV).
Additionally these searches provide strong sensitivity to the production
of a Higgs boson at lower masses. The expected or observed limits 
at $m_H = 125$~GeV are 
3.1 or 3.0 and 3.6 or 4.6 times the SM Higgs boson cross section 
for the CDF and D0 searches, respectively.  
The limits for the different search topologies at $m_H = 125$~GeV
are given in Table~\ref{tab:cdfacc}.

\section{LHC searches in bosonic Higgs boson decays}


The Higgs boson searches at the LHC are performed by decay mode.  In this section we discuss
each search separately, while the combination of
all search results for each experiment is discussed in Sec. IX.  
Previous searches at LEP, Tevatron, and the LHC,
in addition to indirect constraints indicated that the SM Higgs boson had a low mass between approximately
115 and 130 GeV with the region around 125 GeV being of  highest interest.
As of 4 July 2012 the LHC experiments had analyzed searches sensitive to this mass range and higher masses
using datasets of approximately 5 fb$^{-1}$ at 7 TeV and 5.8 (5.3) fb$^{-1}$ for ATLAS (CMS) at 8 TeV.

\subsection{LHC diboson physics}
The key modes for observing the Higgs boson at low mass are the fully reconstructed
$\gamma \gamma$ and $ZZ$  decay modes.  To explore the role of the Higgs
boson in electroweak symmetry breaking the $WW$ decay mode is also crucial.  
Understanding the non-resonant continuum
production of these final states is important to control backgrounds in the Higgs boson
searches.  The LHC experiments have observed $\gamma\gamma$ production in 
7 TeV $pp$ collisions~\cite{Chatrchyan:2011qt}. 
$WW$~\cite{Aad:2011kk,Chatrchyan:2011tz,atlas:2012ks,atlas_ww7,cms_ww7,cms_ww8} 
and $ZZ$~\cite{Aad:2011xj,atlas_zz8,cms_zz7,cms_zz8} production has been observed in
both 7 and 8 TeV collisions and their cross sections measured.  These measurements typically
use identical selections and  techniques to those of the Higgs searches.  Large samples have been collected
in each decay mode.  With future data samples the $WW$ and $ZZ$ modes can be used to study
the electroweak symmetry breaking related phenomena of longitudinal vector boson scattering.  The
contribution of Higgs boson exchange to this process should limit the otherwise divergent behavior
of this process at high energy.

\subsection{Searches in $H\rightarrow \gamma\gamma$}

The LHC experiments have the ability to reconstruct
a Higgs boson in the two photon decay 
mode~\cite{ATLAS:2012ad,atlas_hgg_prel,Chatrchyan:2012tw,cms_hgg_prel}.  
The detectors are designed with the excellent
energy and position resolution necessary to accurately reconstruct
the invariant mass of two photon events.  Excellent mass resolution is
critical since backgrounds from multijet, multijet+photon, and photon+photon
events are large.
This decay has a small branching ratio, since the Higgs boson can
decay only
to photons through a loop diagram involving massive charged
particles. The large inclusive production cross section at lower
masses makes this a viable mode for a Higgs boson observation at low mass.  In addition,
observation of this mode would rule out spin 1 for the observed object.  

Data are collected with diphoton triggers.  
Energy and isolation requirements are made on the photons to
reduce backgrounds.  Converted photons are also used and 
provide good energy and position resolution when electron pairs
can be reconstructed by the trackers.  The ATLAS experiment reduces
background by using the longitudinal segmentation of the calorimeters
to select photons that point to the hard interaction vertex.

Both experiments
apply a separate selection for events with two forward jets to focus on
 a vector
boson fusion like topology which has excellent sensitivity
achieving a signal to background that is an order of magnitude
greater than in topologies without forward jets typical of gluon
fusion production.
In the ATLAS experiment, to optimize sensitivity, events are further divided 
by whether photons are converted
or not and by which pseudorapidity region of the detector
they are reconstructed in,
 since
these classes result in different diphoton mass resolution and as a
result different signal to background.  Finally, events are classified
by the transverse momentum of the diphoton system since backgrounds are
substantially reduced at high values.
In the CMS experiment the output of  a dedicated photon identification BDT,
the transverse momentums of the photons, the opening angle between the photons,
the pseudorapidity of the photons, and the estimated mass resolutions of the diphoton
system are used to classify events by expected signal-to-background ratios using
a BDT multivariate discriminate.  An investigation of the categories
formed by selecting on the BDT output indicates that divisions based on
the transverse momentum of the diphoton system, the detector pseudorapidity
region of the photons, and whether the photons are converted or
unconverted largely determines 
the classification of events. 

\begin{figure}
\includegraphics[width=0.4\textwidth]{{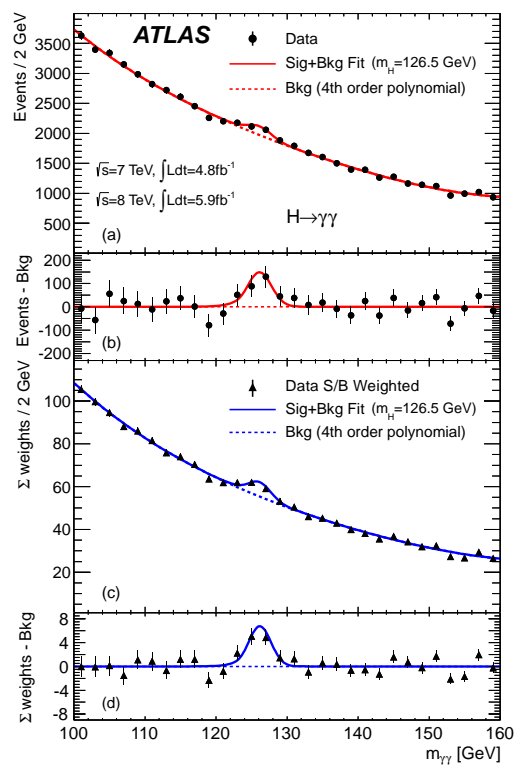}}
\caption{Invariant mass of diphoton events in the ATLAS experiment.
The results are presented with and 
without event weighting by an expected signal-to-background ratio.}
\label{atlas_gaga}
\end{figure}
\begin{figure}
\includegraphics[width=0.36\textwidth]{{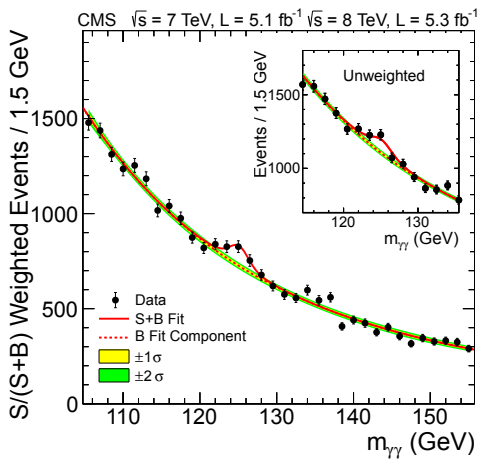}}
\caption{Invariant mass of diphoton events in the CMS experiment with events
weighted by an expected signal-to-background ratio and the unweighted distribution shown as an inset.}
\label{cms_gaga}
\end{figure}

The sensitivities of the individual categories are taken into account when 
calculating the overall sensitivity
of the analysis and when determining exclusions or signal significances.  
The collaborations weight events by expected signal-to-background 
ratio when displaying the diphoton mass distribution to give a visual
representation of the benefit of this classification.

Using the two photon invariant mass distribution to search
for the Higgs boson, regions at both low mass and higher mass have been
excluded by both experiments leaving only a narrow region of mass unexcluded.
The ATLAS experiment excludes the regions
112-122.5 and 132-143 GeV.
This exclusion extends the lower exclusion bound of the LHC searches below the upper exclusion bound from the LEP searches, 
and when combined with other searches excludes the entire  mass range
below 600 GeV except for the
narrow allowed region around 125 GeV.

These analyses have a strong sensitivity to
the production of a low-mass Higgs boson of around 125~GeV with the expectation of
observing approximately 200 events per experiment at that mass.  In that region
both experiments see a significant excess of events.  
The ATLAS diphoton invariant mass distributions 
both weighted by signal-to-background ratio and unweighted are shown in Fig.~\ref{atlas_gaga}.
The corresponding CMS diphoton invariant
mass distributions are shown in Fig.~\ref{cms_gaga}.

The ATLAS experiment sees a $4.5\sigma$ excess of events compatible
with a narrow resonance of mass 126.5 GeV with a signal strength of $1.9 \pm 0.5$ times the SM expectation. 
The CMS experiment sees a $4.1\sigma$ excess of events compatible
with a narrow resonance of mass 125 GeV with a signal strength of $1.6 \pm 0.4$
times the SM expectation.
The strong evidence seen by both experiments in this single decay mode indicates that a new
boson has been observed and strongly disfavors spin one as a possible spin value.
Further characterization of this excess is given next.

\subsection{Searches in 
$H\rightarrow ZZ \rightarrow  \ell^+ \ell^- \ell^+ \ell^-$}

Unique to the LHC experiments is the ability to observe the
Higgs boson over a large range of masses through inclusive production and 
$H\rightarrow ZZ \rightarrow  \ell^+ \ell^-\ell^+ \ell^-$
decay~\cite{ATLAS:2012ac,atlas_hzz_prel,Chatrchyan:2012dg,cms_hzz_prel}.
In the CMS search the charged leptons from $Z$ decay considered at 7 TeV include the $\tau$ leptons~\cite{Chatrchyan:2012hr}.
The LHC experiments are designed to provide large angular coverage for
lepton identification in order  to detect events in this mode at an adequate rate.  Because of the
excellent lepton momentum resolution of the experiments, the Higgs boson
mass can be reconstructed with sufficient precision that background
rates are very low and event counts on order of 10 events are sufficient for
discovery.  With the assumption of SM production rate
the coupling to the $Z$ boson can be measured, and,  by comparison with the $WW$
mode, the ratio of $W$ and $Z$ couplings can be measured.  Finally,
with larger data samples than reported here, the spin and parity of a possible Higgs
boson can be determined solely from this mode using an angular
analysis of the decay products.

Data are collected using single and dilepton (CMS) triggers.  
Transverse momentum requirements are made on the leptons to reduce backgrounds.  
Advanced techniques such as multivariate lepton identification
are applied to maximize lepton finding efficiency.
The CMS experiment improves mass resolution
by using an algorithm designed to detect and recover the
momentum of final state photons radiated by the leptons.
ATLAS incorporates a similar technique as part of its electron momentum fit.
In ATLAS one opposite charge same flavor pair of
leptons is required to be consistent with the $Z$ boson mass.
After these requirements the most significant
background is direct SM $ZZ$ production.  A Higgs boson signal can be
distinguished from the background by looking for a narrow
resonance in the four lepton invariant mass distribution.
For a Higgs boson mass of 125 GeV the ATLAS and CMS experiments
expect $\simeq$  10 events each.

The performance of the four lepton search including selection efficiency
estimates and the scale and resolution of the four lepton invariant mass
is tested by both experiments by searching
for the four lepton final state produced by a $Z$ boson where one initial decay lepton
radiates a photon which converts to a lepton and antilepton pair $Z \rightarrow 4\ell$.
With looser selection on lepton transverse momentum and 
the dilepton masses of same flavor opposite sign pairs this mode can be detected
with substantially higher statistical precision than a Higgs boson at lower mass.
Both experiments perform this analysis within the framework of the
Higgs boson search
analysis and observe this decay with the 
expected performance~\cite{atlas_hzz_prel,cms_hzz_prel,cms_z4l}.

In both experiments the dominant SM $ZZ$ background is estimated from simulation
normalized to NLO cross section predictions. 
In ATLAS the backgrond from $\ell^+\ell^-$+X events, which is dominated by
$Z$+$b\bar{b}$ and $t\bar{t}$ events, is estimated by measuring the
normalization of these backgrounds using selections designed to select
nonprompt  muons or nonisolated electrons, which are more likely to
originate from $b$ jets, and using a transfer factor from simulation
to extrapolate the background prediction to the signal region.
In CMS the backgrounds from $Z$+X events and $t\bar{t}$
are estimated in a control region with one same flavor opposite
charge dilepton pair with an invariant mass consistent with the $Z$ boson and
additional objects.  Using a subset of events from this region with
one additional identified lepton the  probability for an object to be
falsely identified as a charged lepton from $Z$ decay is measured.  That
misidentification rate is applied to determine the number of events with a $Z$ boson and two additional
lepton candidates that are not from $Z$ boson decay in the signal
region.

Both experiments use the four lepton invariant mass distribution to search
for a Higgs boson.  CMS further uses angular information based on the
expected scalar spin zero and parity even nature of the Higgs boson in a matrix
element likelihood analysis(MELA).
Based on the searches large regions at high mass are excluded.
The ATLAS experiment excludes the regions
131-162 and 170-460 GeV, while the CMS
experiment excludes the regions 131-162 and 172-525 GeV.
These are the largest exclusions from a single analysis channel, failing
to exclude only the mass region where the $WW$ branching ratio dominates just above the
on-shell $WW$ production threshold, and in the low-mass region.

\begin{figure}
\includegraphics[width=0.38\textwidth]{{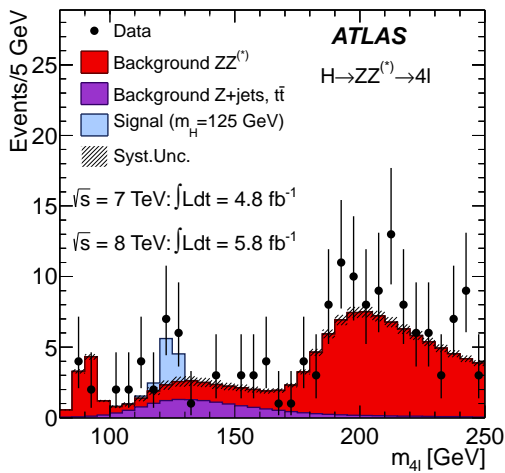}}
\caption{The four lepton invariant mass distribution from the ATLAS experiment.  The
data are displayed as point and the background expectation as a histogram.  Several
SM Higgs boson signal contributions are included for 
different hypothetical Higgs boson masses. Background $Z$+jets and
$\mathrm{t\bar{t}}$ bottom, background $ZZ$ middle and Higgs boson signal top.}
\label{atlas_zz}
\end{figure}

\begin{figure}
\includegraphics[width=0.45\textwidth]{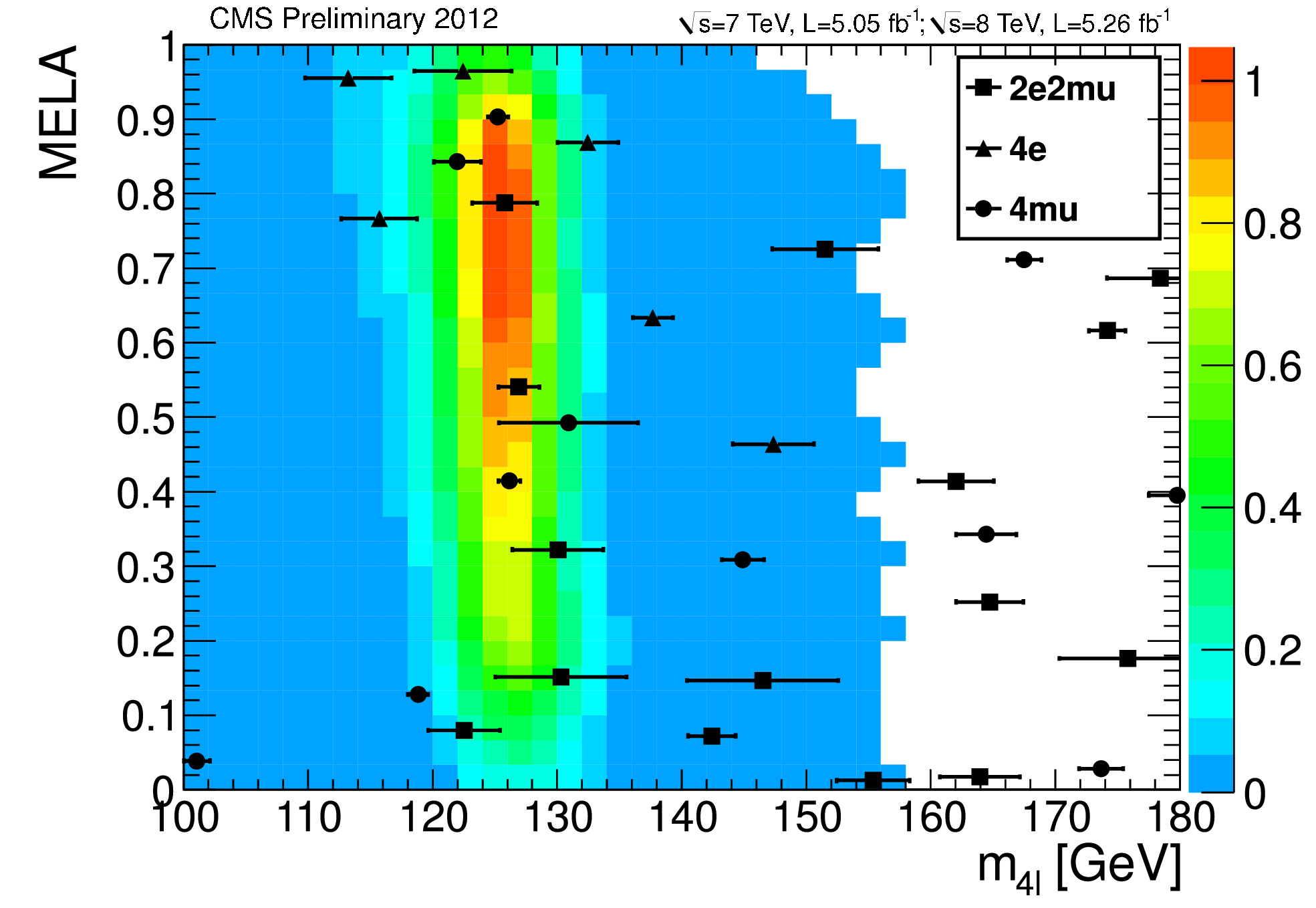}
\caption{A two-dimensional plot of four lepton invariant mass versus
matrix element likelihood from the CMS experiment.  Data are shown with event by event
mass uncertainties while the expectation of a 125 GeV SM Higgs boson is superimposed
as a temperature plot. The central region around 125 GeV is highest in probability.}
\label{cms_zz}
\end{figure}

These analyses have strong sensitivity to the
production of a low-mass Higgs boson of roughly 125 GeV, as can be seen
from the four lepton mass distribution from ATLAS, shown in Fig.~\ref{atlas_zz}, and
from the invariant mass distributions of the four lepton candidates 
versus the MELA discriminant from CMS shown in Fig.~\ref{cms_zz}.

Both experiments see a significant excess of events. 
The ATLAS experiment sees a 3.4$\sigma$ excess of events compatible
with a narrow resonance of mass 125 GeV with a signal strength of 1.3 times the SM expectation. 
The CMS experiment sees a 3.2$\sigma$ excess of events compatible
with a narrow resonance of mass 125.6 GeV with a signal strength of approximately 0.7
times the SM expectation.
The evidence presented by both experiments of a narrow resonance with decays to $ZZ$
indicates that a new boson has been observed, and the use of angular information to
enhance the signal in the CMS case weakly favors spin zero and parity even as quantum
numbers for the new boson though no definitive measurement of these properties
is yet possible.   Further characterization of this excess is given next.

\subsection{Searches in $H\rightarrow W^+W^- \rightarrow \ell^+\nu\ell^-\bar{\nu}$}

The LHC experiments search for inclusive Higgs boson production with the
decay $H\rightarrow W^+W^- \rightarrow \ell^+\nu\ell^-\bar{\nu}$~\cite{Aad:2012sc,atlas_hww_prel,Chatrchyan:2012ty,cms_hww_prel}.
The decay is not fully reconstructed because of the neutrinos
in the final state.  However, the observation of collinear
charged leptons in this decay mode is a distinct signature for the decay of a scalar particle.
Observation of the Higgs boson in this decay mode also excludes spin one as a
potential spin state.
Also, using
the production rate from theory, the mode can be used to determine the
Higgs boson coupling to the $W$ boson or the ratio of $W$ and $Z$
couplings through
comparison with the $H\rightarrow ZZ$ mode.  In addition,
by comparing events with zero or one jet to events with two forward
jets, the gluon fusion and vector boson fusion production rates can be
compared.

The data for the searches are collected using single lepton triggers (ATLAS) and
dilepton triggers (CMS).
Requirements are made on the transverse
momentum of the charged lepton candidates, the magnitude of the
missing transverse energy and on the \MET\ direction, which must not be collinear with
other physics objects in the event.  Also \MET\ measured using
the calorimeter and tracker are compared to require compatibility and
reject events with false \MET .  Loose selection on the collinearity
of the charged leptons is imposed to be consistent with the decay
of a spin zero object to a $W$ boson pair with subsequent leptonic decays.

ATLAS considers only electron and muon events, since such final state
does not have significant Drell-Yan backgrounds and is substantially more sensitive.
CMS divides events into same flavor, electron or muon, and different
flavor, electron and muon, subsamples. 
Both experiments divide the data into events with zero, one, or two or more
jets.  The primary backgrounds are SM direct diboson production, $W$ + jets,
Drell-Yan, single top, and $\ttbar$.  The $\ttbar$ background is dominant and the 
division by jet counting is designed to isolate 
the lower jet categories which have smaller top contributions.  In addition 
$b$-tag vetoes are applied including, at CMS,
the vetoing of jets under a jet transverse momentum threshold.  
The two experiments apply additional selection criteria to further take advantage
of the collinear nature of the charged leptons in Higgs boson decay to $W$ bosons.
In the two jet analysis the jets are required to have a large rapidity
difference and large jet-jet invariant mass to be consistent with the
forward jets from vector boson fusion.  

Both experiments construct orthogonal control regions to study
background kinematics and normalize background contributions.
To study $t\bar{t}$ and $tW$ backgrounds a region is constructed with
no jet multiplicity selection (ATLAS), or requiring
one $b$ tagged jet above threshold (CMS). 
For zero jet events the background is normalized from this region by
extrapolating to the zero jet topology using top event kinematics
from simulation.  For events with jets the background is normalized 
in these regions by applying $b$-tagging and $b$ mistagging
rates measured in a $t \bar{t}$ dominated region to calculate
the number of top events that fail the $b$-tagging criteria in the data.
In addition the region with one $b$-tagged jet can be used to study
the performance of under-threshold $b$ jets to cross-check the performance
of the veto on $b$-tagged under-threshold jets that is used in zero jet
events used in CMS. 
A region with diepton invariant mass larger
than 80 GeV  (ATLAS) or  100 GeV  (CMS) is used to normalize the $WW$
contribution.  
In ATLAS the contribution of Drell-Yan events produced off the $Z$ resonance is
estimated using simulation after additional selection on \MET\ and the
transverse momentum of the dilepton system to reduce this contribution.
In CMS, this background is
estimated by measuring the $Z$ resonance rate, subtracting non-Drell-Yan contributions using  $\mu e$ events, and
then extrapolating the result to the off $Z$ resonance mass range using
the expected distribution of the dilepton mass from simulation.
$W$+jets backgrounds are
estimated using a fully data driven method relying on identifying a sample of
$W$+jets events with a second lepton candidate passing 
a loose
plus antilepton selection (ATLAS), or simply 
a loose selection (CMS),  
and applying false lepton identification
rates measured from data.

\begin{figure}[htb]
\vskip -0.4cm
\hspace{-0.5cm} \includegraphics[width=3.6in]{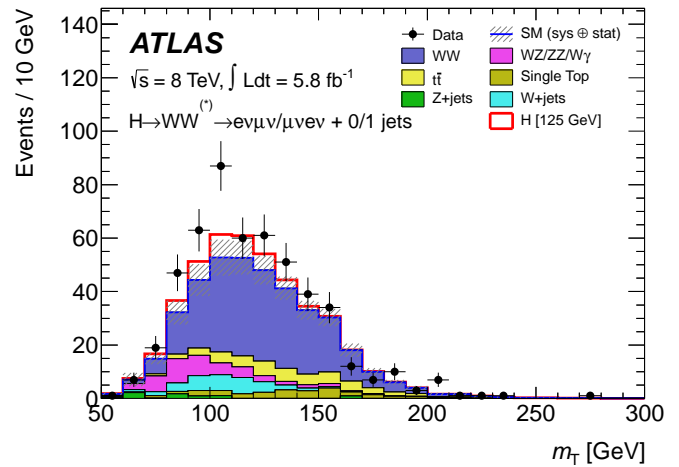}
\caption{
Distribution of the transverse mass in the zero-jet and one-jet $WW$ analyses of ATLAS, with both $e\mu$ and $\mu e$ channels combined. 
The expected signal for $m_H$ = 125 GeV is shown stacked on top of the background prediction. 
}
\label{mtWW-atlas}
\end{figure}

The ATLAS experiment uses the transverse mass of the Higgs boson to search for the signal,
as shown in Fig.~\ref{mtWW-atlas},
while CMS  uses a BDT algorithm
to distinguish signal from background.

These analyses extend the
sensitivity of the searches for pairs of vector bosons
to lower masses including sensitivity to masses as low as 125 GeV.
The experiments exclude the ranges greater than 137 GeV (ATLAS) and 129-520 GeV (CMS).
No upper limit is placed in the ATLAS analysis as the analysis was optimized for
low Higgs boson mass values.
Both experiments have the sensitivity to exclude a Higgs boson at smaller
masses but observe an excess around $\simeq 125$ GeV.
The ATLAS excess has a local significance of 2.8 $\sigma$ and corresponds to a signal strength of
$1.4 \pm 0.5$ times the expected SM rate assuming a Higgs boson mass of 125 GeV.  The CMS experiment
sees a small excess with a local significance of 1.6~$\sigma$.

\subsection{Searches in $H \rightarrow WW,ZZ$ with decays of one boson
to quarks or neutrinos}

The ATLAS and CMS experiments conduct inclusive searches for the Higgs boson
in the decay modes 
$H \rightarrow W^+W^- \rightarrow \ell^+\nu q\bar{q}$ ~\cite{Aad:2012vz,cms_hwwlnuqq_prel,cms_hwwlnuqq_prel2},
$H \rightarrow ZZ \rightarrow \ell^+\ell^- q\bar{q}$~\cite{Aad:2012vw,Chatrchyan:2012sn} and 
$H \rightarrow ZZ \rightarrow \ell^+\ell^- \nu\bar{\nu}$~\cite{atlas:2012va,Chatrchyan:2012ft,cms_hzzllnunu_prel}.
Although these search modes have higher backgrounds than those of 
the other decay modes to pairs of vector bosons 
discussed earlier they have strong sensitivity for high-mass Higgs bosons where
the vector bosons  have high transverse momentum 
leading to significantly reduced backgrounds.

Data are collected on one and two lepton triggers for the analyses with
$W$ and $Z$ bosons, respectively.  The charged leptons are required to pass
transverse momentum and identification requirements.  In the modes with 
$Z\rightarrow \ell^+\ell^-$ decays the invariant mass of 
the charged leptons are required to be
consistent with the $Z$ boson mass while in the $W\rightarrow \ell\nu$
case the transverse mass formed from the \MET\ and the lepton momentum
is required to be consistent with the expected $W$ boson transverse mass.
In the cases where a $Z$ boson decays to neutrinos the \MET\
is required to be large.  The ATLAS experiment further divides this analysis
into high and low Higgs boson mass versions where the \MET\ is required to be larger
in the high-mass version.  Finally, in the cases where the vector bosons
decay to quarks the jet-jet invariant mass is required to be consistent
with the expectation for a $W$ or $Z$ boson.

The primary backgrounds are from SM diboson production, $\ttbar$ production,
and QCD multijet production where mismeasurement of a jet mimics
one of the leptonic signatures.  In all of these background processes
the candidate diboson  pair is not expected to form a mass resonance, and
the individual
vector bosons are not expected to be boosted.   The experiments apply criteria to exploit
these characteristics including requirements on the boost of individual
vector bosons and the opening angle between vector boson decay products.
After selection they search for the Higgs boson using either the
full mass reconstruction ($H \rightarrow ZZ \rightarrow \ell^+\ell^- q\bar{q}$),
transverse mass reconstruction 
($H \rightarrow ZZ \rightarrow \ell^+\ell^- \nu\bar{\nu}$)
or full mass reconstruction applying a $W$ boson mass constraint in the
$H \rightarrow W^+W^- \rightarrow \ell^+\nu q\bar{q}$ mode. 
The experiments search for a Higgs boson in the range 130-600 GeV, with
varying lower thresholds depending on the analysis,
using 4.7-5.0 fb$^{\rm -1}$ of 7 TeV collision data per experiment, 
while the CMS experiment
additionally includes 5.1 fb$^{\rm -1}$ of 8~TeV collision data in the $H \rightarrow W^+W^- \rightarrow \ell^+\nu q\bar{q}$ and
$H \rightarrow ZZ \rightarrow \ell^+\ell^- \nu\bar{\nu}$ modes.
The combined results
provide substantial constraints on the mass of a high-mass Higgs boson excluding masses from
230 to 600 GeV. The exclusion is
dominated by the $H \rightarrow ZZ \rightarrow \ell^+\ell^- \nu\bar{\nu}$ mode although
extended in the lower mass range by the  $H \rightarrow W^+W^- \rightarrow \ell^+\nu q\bar{q}$ search.

\section{LHC searches in fermionic Higgs boson decays}

\subsection{Searches in $H \rightarrow \tau^+\tau^-$}

\begin{figure}[htb]
\vskip -0.4cm
\hspace{-0.6cm} \includegraphics[width=3.0in]{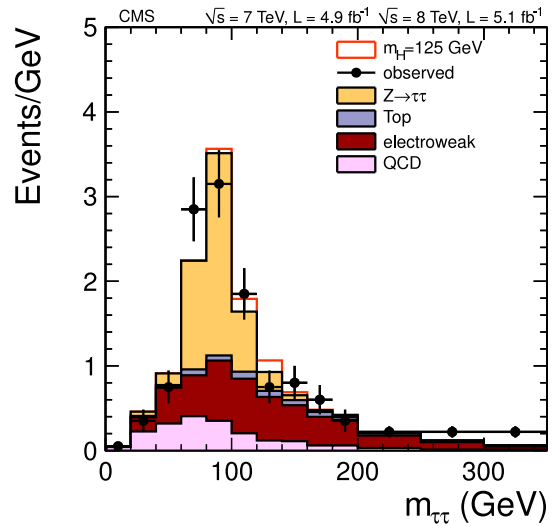}
\caption{
 Distribution of the $\tau\tau$ invariant mass in the combined 7 and 8 TeV data sets 
for the VBF category of the CMS $\tautau$ analysis.
The expected signal for $m_H$ = 125 GeV is shown stacked on top of the background prediction. 
}
\label{mtautau-cms}
\end{figure}

The LHC experiments search for the SM Higgs boson in the 
tau lepton pair decay mode~\cite{Aad:2012ur,Chatrchyan:2012vp,cms_htt_prel}.
The Higgs boson to $\tau$ lepton pair branching ratio is 8\% to 1.5\%
in the Higgs boson mass range of 115-150 GeV, and $\tau$ lepton signals
are distinct enough to make this a viable search mode for all
Higgs boson production processes.  
However, the production mode
with strongest sensitivity is
the vector boson fusion production mode with two associated
forward jets.  This mode is of
high interest since observation of Higgs boson production via
vector boson fusion gives direct information on how the Higgs boson
interacts with high-energy longitudinal vector bosons.  In addition it will
allow the measurement of the coupling of the $\tau$ lepton to the Higgs boson which
should be the first mode to establish a clear signal in a fermionic
decay at the LHC and is the only accessible leptonic coupling in the hadron collider environment,
prior to the planned luminosity upgrades.
The $\tau \tau$ invariant mass obtained in the CMS vector boson analysis is shown in
Fig.~\ref{mtautau-cms}

The experiments search for events with zero, one, or 
two associated jets and oppositely
charged $\tau$ lepton pairs in the following $\tau\tau$ decay final states:
$ee$ (ATLAS), $\mu \mu$, $e \tau_h$, $\mu \tau_h$, and $\tau_h \tau_h$ (ATLAS),
where $e$ indicates $\tau \rightarrow e \nu \nu$, $\mu$ indicates $\tau 
\rightarrow \mu \nu \nu$, and $\tau_h$
where $\tau_h$ indicates
the $\tau$ lepton decayed to hadrons and a $\tau$ neutrino.  
Data are collected on triggers that require one or two charged leptons 
and in the ATLAS experiment one trigger requires two high-$p_T$ $\tau_h$ candidates ($\tau_h \tau_h$).

The primary backgrounds are Drell-Yan production, $W$+jets production, where
one jet is misidentified as a $\tau$ lepton, and $\ttbar$ production.
To enhance the presence of a possible signal relative to backgrounds,
the leptons or sum of hadronic decay products are required to have 
substantial transverse energy and the \MET\ from undetected neutrinos 
is required to be 
collinear with the direction of the dilepton system.
In events with two jets, the two jets are required to
have high momentum and be contained within the forward calorimeters.
There also must be  a large 
rapidity gap between the two jets consistent with the  vector boson fusion hypothesis. 
In events with one jet, the $p_T$ of the jet is required to be high 
to enhance the Higgs boson candidate boost, which improves
separation of the Higgs boson signal from backgrounds
and allows for a more precise estimate of the Higgs
boson mass.

The experiments search for the presence of a Higgs boson using methods designed
to more fully reconstruct the mass of the Higgs boson by including the \MET\
in the calculation and taking advantage of the boosted configurations.  
The ATLAS (CMS) experiments search for a Higgs boson in the range 110-150 GeV
(110-145 GeV) using 4.7-4.9 fb$^{\rm -1}$ of 7 TeV collision data per experiment,
while the CMS experiment
additionally includes 5.1 fb$^{\rm -1}$ of 8~TeV collision data.
The ATLAS experiment reaches
sensitivity to set limits in the range 3-11 times the expected SM production rate, while
the CMS experiment reaches sensitivities of 1.3-2.4 times the SM rate.
At a mass of 125 GeV the CMS experiment sets a limit on the Higgs boson 
production cross section
of a SM Higgs boson of 1.1 times the expected SM rate.  
The sensitivity of this analysis is sufficient 
to achieve evidence for this decay mode using tens of fb$^{\rm -1}$.

\subsection{Searches in $H \rightarrow b\bar{b}$}
The LHC experiments conduct searches
for associated production of Higgs bosons with a $W$ or a $Z$ boson with subsequent
decay of the Higgs boson to a pair of $b$-quarks~\cite{ATLAS:2012zf,Chatrchyan:2012ww,cms_hbb_prel}.  
The Higgs boson to $b$-quark pairs branching ratio is 70\% to 40\%
in the mass range of 115-135 GeV.  However, the $b$-quark decay
signature is not distinct enough to extract the signal from the background
and the leptonic decay signatures of the massive vector bosons produced
in associated production are also necessary to search for the Higgs boson in
this decay mode.
This search is divided
into three subsearches by the decay mode
of the massive vector boson.  The experiments search for $WH \rightarrow \ell\nu b\bar{b}$, 
$ZH \rightarrow \ell^+\ell^- b\bar{b}$, and $ZH \rightarrow \MET b\bar{b}$ 
where the $Z$ decays to neutrinos, ``observed'' as $\MET$.  A charged lepton ($\ell$) refers to electrons and muons of both electric charges.  
Although this set of production and decay processes is less
sensitive than those of many other Higgs boson search modes, it is important
because it can eventually be used to measure the relative couplings of the Higgs
boson to the $W$ and $Z$ bosons and uniquely measure its coupling to $b$ quarks.
 
Data are collected on triggers that require a single charged lepton 
for events with $W$ decays, single leptons (ATLAS $ee$ mode), or pairs of leptons 
for $Z$ decays to charged leptons, and
either \MET\ (ATLAS) or \MET+jets (CMS) for events with $Z$ decays to neutrinos.

Backgrounds such as $W$+jets, single top and $\ttbar$, dibosons, and QCD
multijet production with misidentified leptons are several orders of 
magnitude larger than the signal.  To reconstruct a possible signal the analyses
make several additional requirements.  Leptons must be fully reconstructed by both
the tracking system and dedicated lepton identification systems and pass minimum
transverse momentum thresholds.  Missing transverse momentum must not
be collinear with the jets.  In addition, the ATLAS experiment requires that \MET\
reconstructed using calorimeter and tracking-based algorithms is consistent, while the
CMS experiment uses particle flow-based objects to take advantage of all
detector subsystems. Two jets must be identified as $b$ jets, as expected
from the $b$-quarks from the Higgs boson decay.

Finally in the $ZH \rightarrow \MET b\bar{b}$ mode the \MET\ is required to be
very large.  
The ATLAS experiment
separates the analysis into separate channels for different ranges of
vector boson $p_T$, while
the CMS experiment selects events based on 
the transverse momenta  of both the vector
and Higgs boson candidates. 

The ATLAS experiment uses the
invariant mass of the two $b$ jets to search for Higgs boson candidates 
within each channel, and combines the results into a single search.
The CMS experiment uses variables associated with the above quantities and the 
dijet invariant mass of the Higgs boson as inputs to multivariate discriminants
used to distinguish signal from background.  
The ATLAS (CMS) experiment search for a Higgs boson in the range 110-130 GeV
(100-135 GeV)
using 4.7-5.0 fb$^{\rm -1}$ of 7 TeV collision data per experiment, while the CMS experiment
additionally includes 5.1 fb$^{\rm -1}$ of 8~TeV collision data.  The CMS analyses have sensitivity to set limits 
on the order of 1 to 5 times the expected SM Higgs boson cross section,
depending on the mass, achieving a sensitivity of 1.6 at a Higgs boson mass of 125~GeV.  
The sensitivity of this analysis is be sufficient 
to achieve evidence for this decay mode using tens of fb$^{\rm -1}$.


\subsection{Searches for $t\bar{t}H$  production}

The LHC experiments search for associated production of the Higgs boson
with $t\bar{t}$ where the Higgs boson is radiated from one of the top
quarks.  The strong coupling between the Higgs boson and
the top quark increases the probability of such radiation.
This process provides a direct measurement of the top-quark Yukawa
coupling, which is expected to be one indicating maximal coupling.
Statistically it will not be as significant as measuring the coupling in
gluon fusion events, which are dominated by the top loop diagram, but it will not
suffer from theoretical uncertainties associated with understanding
the gluon fusion loop process.  The ATLAS and CMS experiments have performed
an analysis with the full 7~TeV data set~\cite{atlas_ttH_prel,cms_ttH_prel} in the 
final states with two leptons (CMS only)
and with lepton plus jets decay modes of the top quarks with Higgs boson decay to
$b\bar{b}$.  Events with 4-6 (2-6) jets and 0-4 (2-4) $b$-tagged jets are considered at ATLAS (CMS).
The ATLAS experiment uses the $m_{bb}$ and $H_T$(total energy) distributions to
set limits, while the CMS experiment uses an MVA discriminant.	
The CMS experiment has sensitivity to set limits
on the order of 3 to 9 times the expected SM Higgs boson cross section
achieving a sensitivity of 4.6 at a Higgs boson mass of 125~GeV.  With
a data set on the order of
hundreds of fb$^{\rm -1}$ it is be possible to directly measure
the top quark Yukawa coupling in this channel.

\section{ATLAS, CMS, and Tevatron Results}

        \subsection{Limits and combination methods}

At the LHC
and Tevatron limits are
calculated using the modified frequentist
$CL_{s}$ approach. At the Tevatron,
a Bayesian technique is also used.
The techniques have been shown to produce similar results at the level
of about 5\%.
To facilitate comparisons with the SM and to accommodate
analyses with different degrees of sensitivity and acceptance for more
than one signal production mechanism,
the limits are divided by the SM Higgs boson production
cross section, as a function of Higgs boson mass, for test masses for
which the experiments have performed dedicated searches in different
channels.  A value of the combined limit ratio $\cal{R}$ which is less than or
equal to 1 indicates that that particular Higgs boson mass is excluded
at the 95\% C.L.
Expected limits are calculated both for the background only hypothesis
($B$), for which only SM background contributions are present
in the selected data samples, and for the signal-plus-background
hypothesis. The signal-plus-background hypothesis is calculated by
also including the simulated signal contribution in the limit setting procedure. The
limits are generally determined using the MVA output distributions or the 
invariant mass distributions,
together with their associated uncertainties, as discriminating inputs
to the limit setting procedure.

In the $CL_s$ approach, each hypothesis is tested by simulating the outcome of multiple
pseudoexperiments. The data are assumed to be drawn from a Poisson
statistical parent distribution, and each pseudo experiment result is
obtained by randomly generating pseudodata using a Poisson distribution
for which the mean is taken from either the background-only or signal-plus-background 
hypothesis.  To evaluate the statistical significance of each
result a negative Poisson log-likelihood ratio (LLR) test statistic is
evaluated, and the outcomes are ordered in terms of their contributing
statistical significance. The frequency of each outcome is used to define
the shape of the resulting LLR distributions at each mass point for both
the background-only and signal-plus-background hypotheses.

Systematic uncertainties in each hypothesis are accounted for by  nuisance parameters which
are assigned an a prior probability distribution. These parameters refer
to uncertainties in the expected background contributions and, in the
case of the signal-plus-background hypothesis, 
also uncertainties on the simulated signal contribution.
The nuisance parameters are Gaussian and randomly assigned within the parent
distribution for each pseudo experiment. Correlations between the uncertainties
are taken into account.
To minimize
the impact of the nuisance parameters the profile likelihood distribution is
maximized over the nuisance parameters within each pseudo experiment, once for
the background-only and once for the signal-plus-background hypotheses.
Each background is allowed to vary within its uncertainties by varying the
nuisance parameters in the fitting procedure, while the fit is constrained to lie
within the uncertainties.

The expected limits are calculated with respect to the median of the
background-only LLR distribution, whereas the observed limits are quoted with respect to the
single LLR value of the actual measurement.  The distribution of expected limits can also
be analyzed to understand 1$\sigma$ and 2$\sigma$ deviations from median.

This framework can also produce statistical results quantifying the expectation for
and properties of a signal.  Given the SM expectation for signal contributions, the expected $p$-value or probability for backgrounds to fluctuate
to the statistical significance of the expected signal can be computed.
Similarly, given an excess in the data,
the observed $p$-value can be computed.  Finally, in this technique
the SM Higgs signal cross section is multiplied by an arbitrary factor that is
fit for the likelihood minimization allowing for a measurement of the observed
cross section.

\begin{table*}[hbt]
\begin{center}
\caption{The most significant excesses seen in ATLAS and CMS results
  and the combined local and global significances or
P values.}
\label{table:lhcsignal}
\begin{small}
\begin{tabular}{|c|c|c|}\hline
Topology & ATLAS Significance and Mass & CMS Significance and Mass \\ \hline
$H\rightarrow WW \rightarrow \ell\nu\ell{\nu}$     & 2.8~$\sigma$ \@  125.0~GeV   &  1.6~$\sigma$ \@  125.0~GeV    \\
$ H \rightarrow ZZ \rightarrow 4\ell$     & 3.4~$\sigma$ \@  125.0~GeV   & 3.1~$\sigma$ \@  125.6~GeV    \\
$ H  \rightarrow \gamma\gamma$                               & 4.5~$\sigma$ \@  126.5~GeV & 4.1~$\sigma$ \@  125.0~GeV  \\ 
\hline
Combined Significance   & 5.9~$\sigma$ \@  126.0~GeV& 5.0~$\sigma$ \@  125.3~GeV \\
\hline
\end{tabular}
\end{small}
\end{center}
\end{table*}

\begin{center}
\begin{table*}[t]
\caption{\label{tab:lhcsearches}The integrated luminosity, explored mass range, 95\% C.L. expected and observed limits,
and references for the ATLAS and CMS analyses.
For analyses without SM sensitivity the expected and observed exclusions on cross section
normalized to the SM expectation ($\cal{R}$) assuming $m_H=125$ are given.  For analyses with
SM sensitivity expected and excluded ranges of mass are given.
}
\begin{ruledtabular}
\begin{tabular}{lccccccc} \\
ATLAS Channels & Luminosity  & $m_H$ range & Expected exclusion & Observed exclusion   & Reference \\
        & (7+8 TeV, fb$^{-1}$) & (GeV) &  (Range in GeV, $\cal{R}$) & (Range in GeV, $\cal{R}$) &          \\ \hline
$t{t}H \rightarrow t{t} b{b}$  & 4.7      & 110-130 & --               & --                 &\cite{atlas_ttH_prel} \\
$VH\rightarrow Vb\bar{b}$                   & 4.7      & 110-130 & $\cal{R}=$4.0              & $\cal{R}=$4.6                &\cite{ATLAS:2012zf} \\
$H  \rightarrow \tau^+ \tau^-$             & 4.7      & 100-150 & $\cal{R}=$3.3              & $\cal{R}=$3.4          &\cite{Aad:2012ur} \\
$H \rightarrow \gamma \gamma$              & 4.8+5.9  & 110-150 & 110-139.5        & 112-122.5 \& 132-143 &\cite{ATLAS:2012ad,atlas_hgg_prel} \\
$H\rightarrow WW \rightarrow \ell\nu\ell\nu$   & 4.7+5.8  & 110-600 & $>$124.5           & $>$137.0             &\cite{Aad:2012sc,atlas_hww_prel} \\
$H\rightarrow WW \rightarrow \ell\nu 2$jet & 4.7      & 300-600 & --               & --                 &\cite{Aad:2012vz} \\
$H\rightarrow ZZ \rightarrow 4  \ell $     & 4.8+5.8  & 110-600 & 124-164 \& 176-500 & 131-162 \& 170-460   &\cite{ATLAS:2012ac,atlas_hzz_prel} \\
$H\rightarrow ZZ \rightarrow 2  \ell 2 $jet& 4.7      & 200-600 & 351-404          & 300-322 \& 353-410    &\cite{Aad:2012vw} \\
$H\rightarrow ZZ \rightarrow 2  \ell 2 \nu$& 4.7      & 200-600 & 280-497          & 319-558            &\cite{atlas:2012va} \\
\hline
ATLAS Combined                            & 4.7+5.8  & 110-600 & 110-582          & 111-122 \& 131-559   &\cite{ATLAS:2012gk} \\
\hline
\hline\\
  CMS Channels & Luminosity  & $m_H$ range & Expected exclusion & Observed exclusion   & Reference \\
        & (7+8 TeV, fb$^{-1}$) & (GeV) &  (range in GeV, $\cal{R}$) & (range in GeV, $\cal{R}$) &          \\ \hline
$t{t}H \rightarrow t{t} b{b}$  & 5.0      & 110-130 & $\cal{R}=$4.6          & $\cal{R}=$3.8               &\cite{cms_ttH_prel} \\
$VH\rightarrow Vb\bar{b}$                   & 5.0+5.1  & 110-135 & $\cal{R}=$1.6         & $\cal{R}=$2.1              &\cite{Chatrchyan:2012ww,cms_hbb_prel} \\
$H  \rightarrow \tau^+ \tau^-$             & 4.9+5.1  & 110-145 & $\cal{R}=$1.28         & $\cal{R}=$1.06              &\cite{Chatrchyan:2012vp,cms_htt_prel} \\
$H \rightarrow \gamma \gamma$              & 5.1+5.3  & 110-150 & 110-144      & --                &\cite{Chatrchyan:2012tw,cms_hgg_prel} \\
$H\rightarrow WW \rightarrow \ell\nu\ell\nu$   & 4.9+5.0  & 120-600 & 122-450      & 129-520           &\cite{Chatrchyan:2012ty,cms_hww_prel} \\
$H\rightarrow WW \rightarrow \ell\nu 2$jet & 5.0+5.1  & 170-600 & 220-515 & 230-480        &\cite{cms_hwwlnuqq_prel,cms_hwwlnuqq_prel2} \\
$H\rightarrow ZZ \rightarrow 4  \ell $     & 5.1+5.3  & 110-600 & 121-570      &131-162 \& 172-525  &\cite{Chatrchyan:2012dg,cms_hzz_prel} \\
$H\rightarrow ZZ \rightarrow 2  \ell 2\tau$& 4.6      & 190-600 & --           & --               &\cite{Chatrchyan:2012hr} \\

$H\rightarrow ZZ \rightarrow 2  \ell 2 $jet& 4.6      & 130-600 & --           & --               &\cite{Chatrchyan:2012sn} \\
$H\rightarrow ZZ \rightarrow 2  \ell 2 \nu$& 5.0+5.0  & 200-600 & 290-530      & 278-600          &\cite{Chatrchyan:2012ft,cms_hzzllnunu_prel} \\
\hline
CMS Combined                              & 5.1+5.3  & 110-600 & 110-600      & 110-122.5 \& 127-600 &\cite{CMS:2012gu} \\

\end{tabular}
\end{ruledtabular}
\end{table*}
\end{center}

\subsection{ATLAS and CMS results}
The ATLAS and CMS experiments analyze their data using the statistical
techniques described previously.  Each analysis channel is analyzed separately
and within each experiment the Higgs boson search results are 
combined~\cite{ATLAS:2012ae,atlas_comb_prel,ATLAS:2012gk,Chatrchyan:2012tx,CMS:2012gu}.  
Table~\ref{tab:lhcsearches} summarizes for ATLAS and CMS,
the integrated luminosities, the Higgs boson mass ranges over which the searches are performed,
and references to further details for each analysis.  
Also given are expected and observed
exclusion ranges for SM Higgs boson production at 95\% C.L. or for those channels which do not have sensitivity
to limit the SM rate of Higgs boson production at any mass, the expected and observed limits on
cross section for the SM Higgs boson with mass $m_H = 125$ expressed as a multiplicative
factor times the predicted SM rate.
Depending on the mass of a hypothetical Higgs boson, the
LHC experiments have the sensitivity to discover the Higgs boson in individual
production and decay channels.  At some masses it is possible to have 
observations of the Higgs boson in several channels.  
The combination of these results allows the experiments to achieve larger
exclusion ranges over masses where no evidence for a Higgs boson signal is
seen, earlier discoveries in mass ranges where several analysis channels have sensitivity, and
comparison among channels to demonstrate the consistency of a possible signal with
a Higgs boson hypothesis.  The LHC experiments are prepared for, but have not yet produced 
a joint combination of these results.  For the first discovery the combination of results from multiple
experiments is not preferred since simultaneous observation constitutes
both an observation and an independent confirmation of the result.

\begin{figure}
\includegraphics[width=0.52\textwidth]{{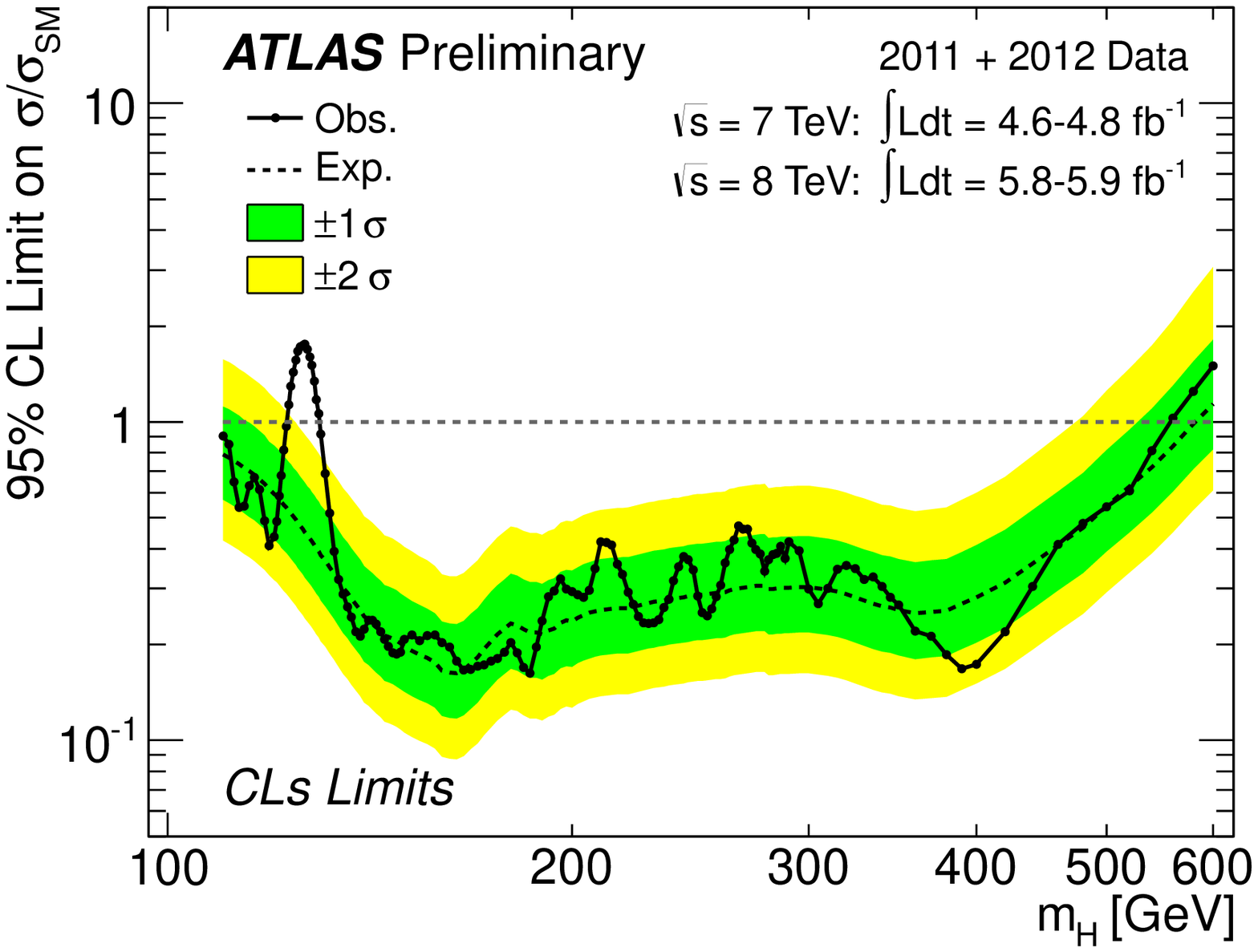}}
\caption{The ATLAS experiments combined upper limit as a function of the
Higgs boson mass between 100 and 600~GeV Solid black: observed limit/SM; dashed black: median expected limit/SM in the background-only hypothesis:
colored bands: $\pm 1,2 \sigma$ distributions around the median expected limit.}
\label{atlas_limits}
\end{figure}

\begin{figure}
\includegraphics[width=0.5\textwidth,height=0.3\textheight]{{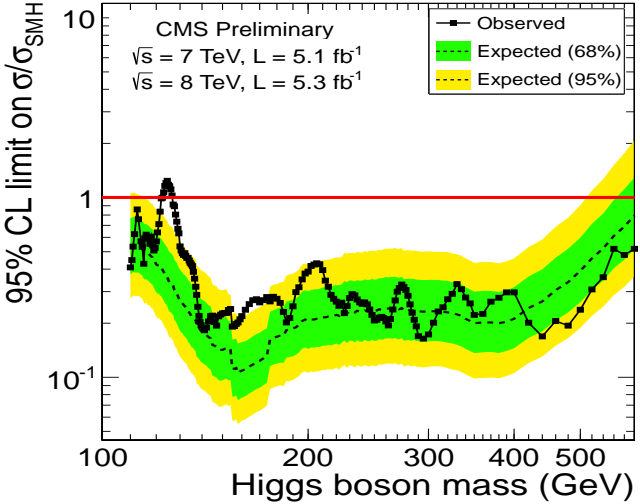}}
\caption{The CMS experiments combined upper limit as a function of the 
Higgs boson mass between 100 and 600~GeV Solid black: observed limit/SM; dashed black: CMS expected limit/SM in the background-only hypothesis:
colored bands: $\pm 1,2 \sigma$ distributions around the median expected limit.}
\label{cms_limits}
\end{figure}

Over a large region of masses the LHC experiments observe no evidence for a Higgs boson.  
The LHC data show a consistent picture with a high-mass SM Higgs boson
typically excluded by multiple channels.
At high mass the ATLAS experiment excludes the production of a SM Higgs boson with masses from 131 to 559 GeV at 95\% C.L. 
and the CMS experiment excludes a region from 128 to 600 GeV at 95\% C.L., where 600 GeV is the limit of the
search range.  At low mass the ATLAS experiment excludes the production of a SM Higgs boson 
with masses from 111 to 122 GeV at 95\% C.L., while
the CMS experiment excludes the region from 110 to 122.5 GeV.
The combined ATLAS and CMS limits are presented in Figs.~\ref{atlas_limits} and~\ref{cms_limits},
respectively.


\begin{figure}[htb]
\includegraphics[width=0.45\textwidth,height=0.26\textheight]{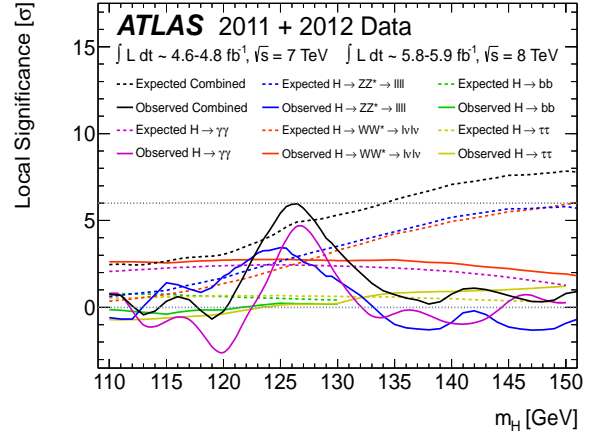} 
\caption{	
ATLAS local significance~\cite{ATLAS:2012gk} for each search channel and the combination.  
The observed significance are shown with solid curves, 
and the median expected significance assuming a signal is present at the SM strength
are shown with dashed curves.   A dash dotted line indicates the $6~\sigma$
threshold.  The highest local significances of the $ZZ$ and $\gamma\gamma$
channels are $3.4\sigma$ and $4.5\sigma$ respectively while the
combined significance of all channels is $5.9\sigma$.
}
\label{atlaspvalue}
\end{figure}

\begin{figure}[htb]
\includegraphics[width=0.45\textwidth,height=0.26\textheight]{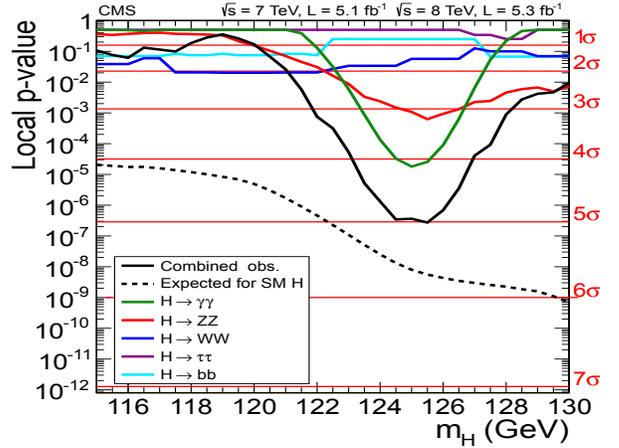} 
\caption{
CMS local $p$ values~\cite{CMS:2012gu}.
The observed
$p$ values are shown with solid curves, and the median expected $p$ value for the combined search
assuming a signal is present at the SM strength
is shown with a dashed curve.  Horizontal lines indicate the $1~\sigma - 7~\sigma$ thresholds.  
The highest local significances of the $ZZ$ and $\gamma\gamma$
channels are $3.1\sigma$ and $4.1\sigma$ respectively while the
combined significance of all channels is $5.1\sigma$.
}
\label{cmspvalue}
\end{figure}

At low masses the experiments have the sensitivity to exclude or observe the Higgs boson.  The sensitivities
for observation of a signal are quantified as an expected $p$-value for the background to 
fluctuate to a signal as large as the median
expectation for a SM Higgs boson.
The combined expected $p$-value at $m_H = ~125$ GeV is $4.9~\sigma$ for the ATLAS experiment and
$5.8~\sigma$ for the CMS experiment.
In the region around 125 GeV both experiments observe an excess of events in multiple search channels.  
The experiments evaluate the $p$ values for each channel separately and for the entire combination and compare
those values with the expected background-only $p$ values given a SM Higgs boson as a function of mass
(see Fig.~\ref{atlaspvalue},\ref{cmspvalue}). 
Information quantifying the most significant excesses in the individual search channels was given previously in the 
sections describing the different LHC Higgs boson searches and is summarized along with the most significant
combined excess from each experiment in Table~\ref{table:lhcsignal}.
Both experiments observe a Higgs boson signal with local significances above the evidence
level of $3~\sigma$ in the $ZZ$ and $\gamma \gamma$ decay modes and combined significances of 
$5.9~\sigma$ at $m_H=126$ GeV for the ATLAS experiment and $5.0~\sigma$ at $m_H=125.3$ for the CMS experiment.
The ATLAS experiment also evaluates a global significance over their entire search range assuming no a
priori knowledge of the SM Higgs boson and finds a global significance of $5.1~\sigma$.  The simultaneous
observation of a new particle with mass of approximately 125 GeV constitutes a definitive discovery.  
The decay modes in which the particle is strongly observed also indicate that the particle is a boson and plays
a role in the mechanism of electroweak symmetry breaking.


\begin{figure}[htb]
\includegraphics[width=0.42\textwidth,height=0.30\textheight]{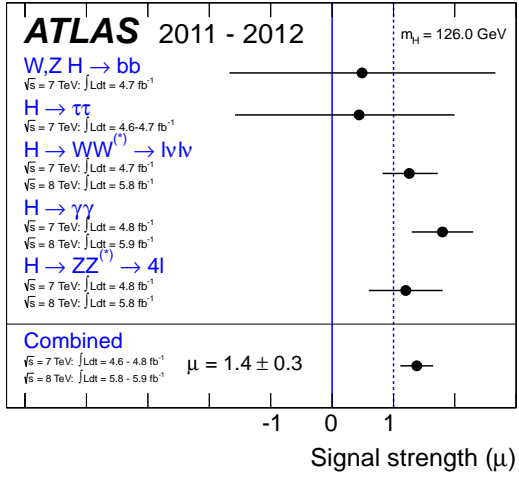} 
\caption{
ATLAS best-fit signal strength for all SM Higgs boson decays for $m_H=125$~GeV/$c^2$. 
}
\label{atlassmxsfit}
\end{figure}
\begin{figure}[htb]
\includegraphics[width=0.50\textwidth,height=0.32\textheight]{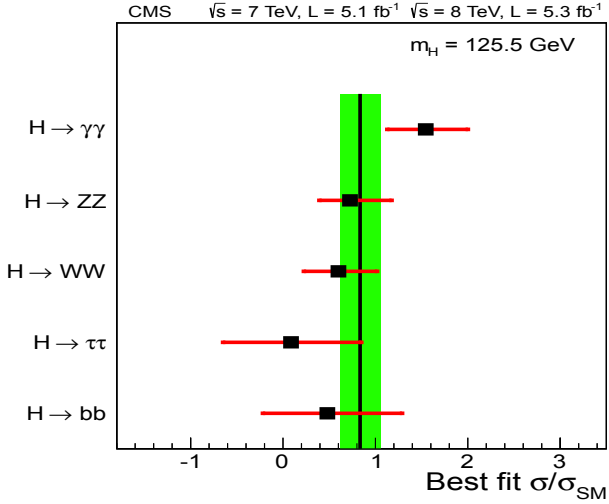} 
\caption{
CMS best-fit signal strength for all SM Higgs  boson decays for $m_H=125$~GeV/$c^2$.
Untagged refers
to the cross section extracted from topologies sensitive to gluon
fusion production.
The shaded band corresponds to the $\pm$ 1$\sigma$ uncertainty on the full combination.
}
\label{cmssmxsfit}
\end{figure}

The CMS and ATLAS Collaborations measured several properties to understand the compatibility
of the observed boson with the SM Higgs boson and present the results in their papers
reporting the observations~\cite{ATLAS:2012gk,CMS:2012gu}.

The experiments fit for the cross section for Higgs boson production given the observed data in
each decay channel and globally combining all decay channels.  The results are presented as a ratio
to the expected SM values in Figs.~\ref{atlassmxsfit} and~\ref{cmssmxsfit} for the ATLAS
and CMS experiments, respectively.  Of note are the larger than expected cross-section times
branching ratios seen in the 
$\gamma \gamma$ (ATLAS and CMS) and $ZZ$ (ATLAS)
decay modes.  
These modes are dominated by gluon fusion production.
The combined signal strengths measured by the experiments are $1.4\pm 0.3$ for ATLAS and $0.87\pm 0.23$ for CMS compatible
with the SM Higgs boson expectation. Individual signal strengths in
the most sensitive modes were discussed in the sections on individual
searches.

\begin{figure}[htb]
\includegraphics[width=0.47\textwidth,height=0.29\textheight]{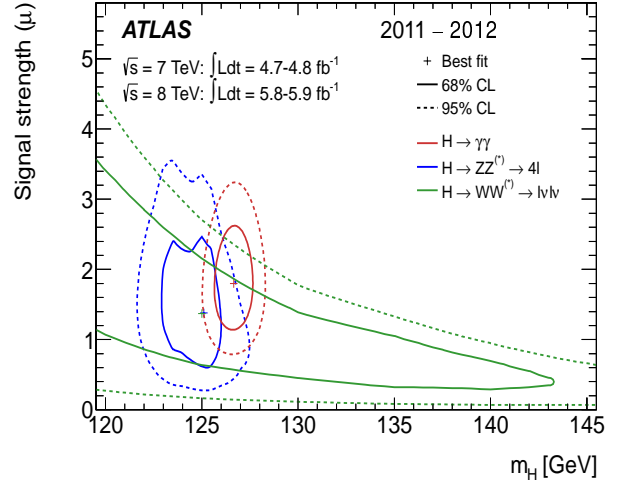} 
\caption{
The ATLAS two dimensional fit for the cross section and compared to the SM expectation and
the Higgs boson mass for highest significance decay channels.
}
\label{atlas_mass}
\end{figure}
\begin{figure}[htb]
\includegraphics[width=0.43\textwidth]{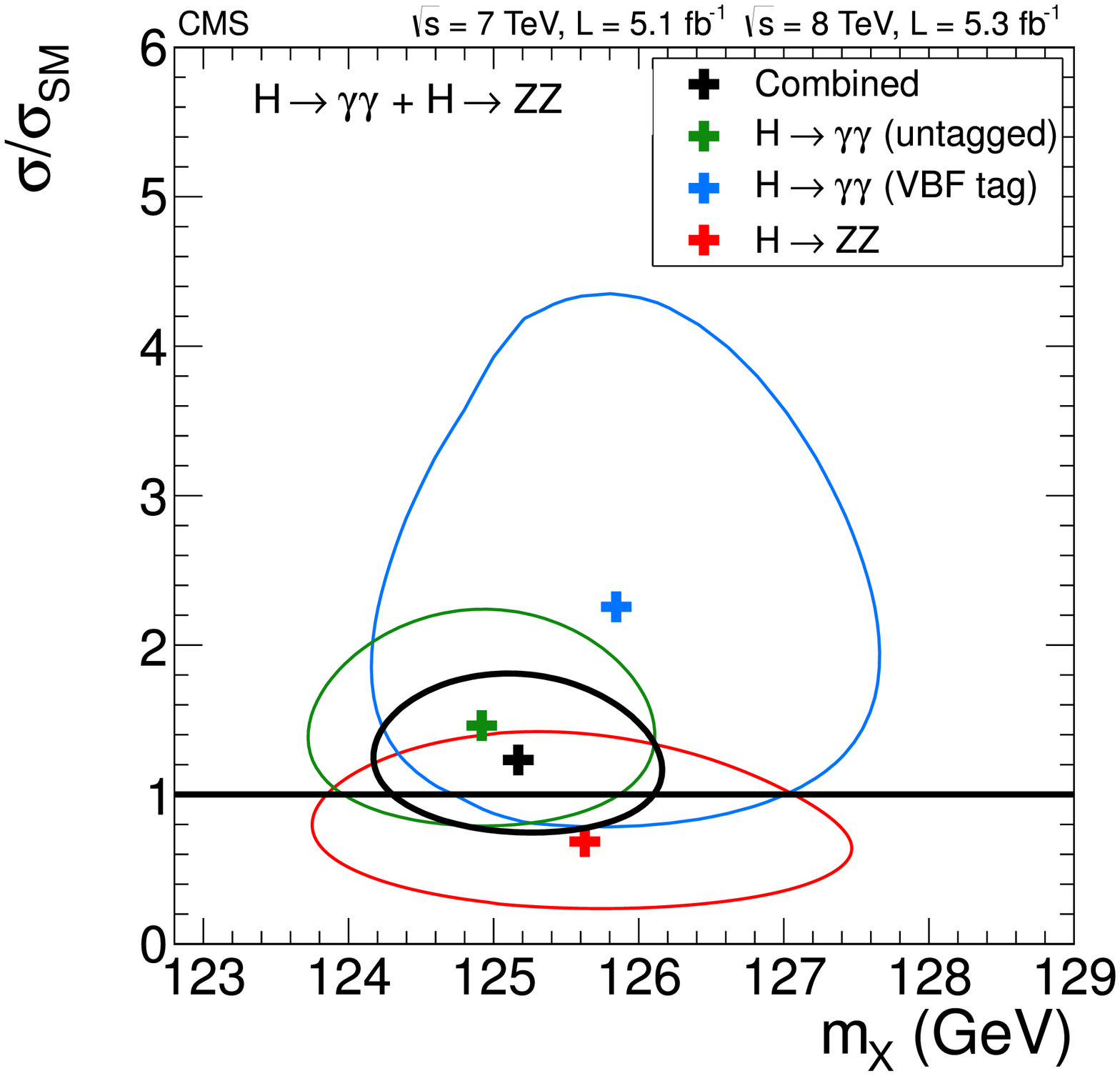} 
\caption{
The CMS two dimensional fit for the cross section and compared to the SM expectation and
the Higgs boson mass for highest significance decay channels and the combined fit using
those channels.
}
\label{cms_mass}
\end{figure}

The fully reconstructed decays of the Higgs boson 
 $ H  \rightarrow \gamma\gamma$  and $H \rightarrow ZZ \rightarrow \ell^+\ell^-\ell^+\ell^-$
have excellent mass resolution.
The  $H\rightarrow W^+W^- \rightarrow \ell^+\nu\ell^-\bar{\nu}$ decay mode has substantial rate but has poor mass resolution due to 
the two neutrinos in the final state.  
The ATLAS experiment measures
a mass for the observed boson of $m_H = 126.0 \pm 0.4(stat) \pm 0.4(sys)$~GeV using all three decay modes.
The individual fits in a two-dimensional analysis of signal strength versus mass are shown in Fig.~\ref{atlas_mass}.
The CMS experiment uses the  fully reconstructed  $ H  \rightarrow \gamma\gamma$  and
$H \rightarrow ZZ \rightarrow \ell^+\ell^-\ell^+\ell^-$ 
modes to measure a mass of $m = 125.3 \pm 0.4(stat) \pm 0.5(sys)$~GeV.
The individual and combined fits are shown in Fig.~\ref{cms_mass}.  The results are compatible with
limits from previous searches and the prediction of the SM Higgs boson mass from constraints 
derived from electroweak measurements.

If the observed boson is involved in the mechanism of electroweak symmetry breaking,
the measurement of its coupling to the $W$ and $Z$ bosons is a crucial discriminant.  The production
and decay rates measured by the experiments are compatible with the SM.
The ratio
of the $W$ and $Z$ couplings 
can be computed by dividing the production times decay rates for 
$H \rightarrow WW$ and 
$H \rightarrow ZZ$ 
since the
production of the Higgs boson takes place via the same mechanisms.
The ATLAS experiment measures $R_{WZ} = 1.07^{+0.35}_{-0.27}$~\cite{atlas_prop_prel} and the CMS experiment measures 
$R_{WZ} = 0.9^{+1.1}_{-0.6}$ 
consistent with the SM expectation where both experiments have normalized the measurement so that the expected
value in the SM is 1.

In summary, the LHC experiments extended the LEP exclusion to 122.5 GeV and further excluded
a SM Higgs boson with mass 
between 128 and 600 GeV.  The ATLAS and CMS experiments both observe a significant excess of events in
the region around 125 GeV with evidence for the production of a new boson
in the $ZZ$ and $\gamma \gamma$ decay modes, with observed local significances of 4.5 $\sigma$ and 4.1 $\sigma$
in the $\gamma \gamma$ mode and 3.4 $\sigma$ and 3.1 $\sigma$ in the $ZZ$ mode.
Significant signals (2.8 $\sigma$ and 1.6 $\sigma$) are also observed
in the $H \rightarrow WW$ decay mode, while the observed significance in the fermionic modes
($H \rightarrow \tau\tau$ and $H \rightarrow b\bar{b}$) is weak, which is not unexpected given
the currently low expected significance in these modes.
When combining all their channels, both experiments independently report the
discovery of a new boson and provide first measurements of its fundamental properties,
in agreement with those expected from a SM Higgs boson with a mass close to 125 GeV.

	\subsection{Tevatron combined results}

As in the LHC experiments, to simplify the combination, the searches are separated into 
mutually exclusive final states.
Table~\ref{tab:cdfacc} summarizes for each CDF and D0 search
the integrated luminosities, the Higgs boson mass ranges over which the searches are performed,
the ratios of expected and observed limits with respect to SM Higgs
boson expectations achieved for $m_H=125$ GeV, and the references to further details for each analysis.
Using the combination procedure outlined in Sec. IX B, 
limits on SM Higgs boson production $\sigma \times B(H\rightarrow X)$
in \pp~collisions at $\sqrt{s}=1.96$~TeV for $100\leq m_H \leq 200$ GeV
are extracted.

The combinations of results from each single 
experiment~\cite{CDFHiggs,DZHiggs}, 
as used in this Tevatron combination, yield the following
ratios of 95\% C.L. observed (expected) limits to the SM expectation:
2.4~(1.2) for CDF and 2.2~(1.6) for D0 at $m_{H}=115$~GeV,
2.9~(1.4) for CDF and 2.5~(1.9) for D0 at $m_{H}=125$~GeV, and
0.42~(0.69) for CDF and 0.94~(0.76) for D0 at $m_{H}=165$~GeV.

\begin{figure}[htb]
\begin{centering}
\includegraphics[width=8.5cm]{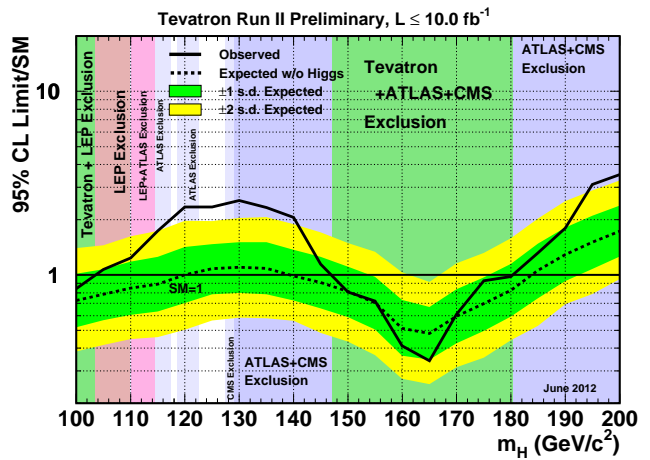}
\caption{
\label{comboRatio}
Observed and expected 
95\% C.L. upper limits on the ratios to the SM cross section, as
functions of the Higgs boson mass
for the combined CDF and D0 analyses.
  The bands indicate the
68\% and 95\% probability regions where the limits can
fluctuate, in the absence of signal.
}
\end{centering}
\end{figure}

The ratios of the 95\% C.L. expected and observed limits to the SM cross
section are shown in Fig.~\ref{comboRatio} for the combined CDF
and D0 analyses.  
The observed (expected) limit values are
1.8~(0.94) at $m_{H}=115$~GeV,
2.2~(1.1) at $m_{H}=125$~GeV,
and
0.39~(0.49) at $m_{H}=165$~GeV.

\begin{center}
\begin{table*}[htb]
\caption{\label{tab:cdfacc}The integrated luminosity, explored mass range, 95\% C.L. expected and observed limits on Higgs boson production cross section relative to
the SM expectation ($\cal{R}$) assuming $m_H=125$ GeV, 
 and references
for the different CDF and D0 analyses grouped by the 
final states ($\ell$ = $e$ or $\mu$) considered.
}
\begin{ruledtabular}
\begin{tabular}{lccccc} \\
CDF Channels & Luminosity (fb$^{\rm -1}$)  & $m_H$ range & exp. $\cal{R}$ & obs. $\cal{R}$    & Reference \\
        & 2 TeV  & (GeV) &  125 GeV & 125 GeV &          \\ \hline
$WH\rightarrow \ell\nu b\bar{b}$ \ \ \ \ \  2-jet \& 3jet channels                  & 9.5 &100$-$150 & 2.8 & 4.9 & \cite{CDFWHprel} \\
$ZH\rightarrow \ell^+\ell^- b\bar{b}$ \ \ \ 2-jet \& 3 jet channels                  & 9.5 &100$-$150 & 3.6 & 7.2 & \cite{CDFZHprel} \\
$ZH\rightarrow \nu\bar{\nu} b\bar{b}$ \ \ \ \ \ \ 2-jet \& 3jet channels             & 9.5 &100$-$150 & 3.6 & 6.8 & \cite{CDFvvbbprel}\\
$H\rightarrow W^+ W^-$  \& $WH\rightarrow WW^+ W^-$ \& $ZH\rightarrow ZW^+ W^-$      & 9.7 &110$-$200 & 3.1 & 3.0 & \cite{cdfHWW} \\
$H \rightarrow \gamma \gamma$ \ \ \                                                  & 10.0&100$-$150 & 9.9  & 17.0 & \cite{cdfHgg} \\
$H\rightarrow ZZ$ \ \ \ (four leptons, limits are given at 130 GeV)              & 9.7 &120$-$200 & 18.3 & 20.5 & \cite{cdfHZZ} \\
$H\rightarrow W^+ W^-$ ($e\tau_{{h}}$)+($\mu\tau_{{h}}$) 
\& $WH\rightarrow WW^+ W^-$ (1 $\tau_{{h}}$)                                         & 9.7 &130$-$200 & . & . & \cite{cdfHWW2} \\
$H$ + $X\rightarrow \tau^+ \tau^-$ \ \ \ (1 jet)+(2 jet)                             & 8.3 &100$-$150 & 14.8 & 11.7 & \cite{cdfHtt} \\
$WH \rightarrow \ell \nu \tau^+ \tau^-$/$ZH \rightarrow \ell^+ \ell^- \tau^+ \tau^-$ & 6.2 &100$-$150 & 23.3 & 26.5 & \cite{cdfVHtt} \\
$WH+ZH\rightarrow jjb{\bar{b}}$ \ \ \  (SS,SJ)                                       & 9.5 &100$-$150 & 9.0  & 11.0 & \cite{cdfjjbb} \\
$t\bar{t}H \rightarrow W W b\bar{b} b\bar{b}$ \ \ \ (no lepton) - (lepton) 
                                                                                 & 5.7-9.5 &100$-$150 & 12.4 & 17.6 & \cite{cdfttHnoLep,cdfttHLep} \\
\hline
\hline\\
D0 Channels & Luminosity (fb$^{\rm -1}$)  & $m_H$ range & exp. $\cal{R}$ & obs. $\cal{R}$    & Reference \\
        & 2 TeV & (GeV) &  125 GeV & 125 GeV &          \\ \hline
$WH\rightarrow \ell\nu b\bar{b}$ \ \ \ \ \   2-jet \& 3jet channels             & 9.7 &100$-$150 &  4.7 &  5.2 & \cite{D0WHprel} \\
$ZH\rightarrow \ell^+\ell^- b\bar{b}$ \ \ \  2-jet \& 3jet channels        & 9.7 &100$-$150 &  5.1 &  7.1 & \cite{D0ZHprel} \\
$ZH\rightarrow \nu\bar{\nu} b\bar{b}$ \ \ \ \ \ \  2-jet channel           & 9.5 &100$-$150 &  3.9 &  4.3 & \cite{D0vvbbprel} \\
$H\rightarrow W^+ W^- \rightarrow \ell^\pm\nu \ell^\mp\nu$ \ \ \       & 8.6-9.7 &115$-$200 &  3.6 &  4.6 & \cite{dzHWW}\\
$H \rightarrow \gamma \gamma$                                              & 9.7 &100$-$150 &  8.2 & 12.9 & \cite{dzHgg} \\
$H\rightarrow W^+ W^- \rightarrow \mu\nu \tau_{{h}}\nu$ \ \ \              & 7.3 &115$-$200 & 12.8 & 15.7 & \cite{dzVHt2}\\
$H\rightarrow W^+ W^- \rightarrow \ell\bar{\nu} jj$                        & 5.4 &130$-$200 &   .  &   .  & \cite{dzHWWjj}\\
$H$+$X$$\rightarrow$$ \ell^\pm \tau^{\mp}_{{h}}jj$  \ \ \              & 4.3-6.2 &105$-$200 & 40.0 &  44.0  & \cite{dzVHt2} \\
$VH \rightarrow \tau \tau \mu + X $ 				           & 7.0 &115$-$200 & 17.6 & 13.1 & \cite{dzttl} \\
$VH \rightarrow e^\pm \mu^\pm + X $ \ \ \                                  & 9.7 &115$-$200 & 11.6 &  7.8 & \cite{dzWWW2} \\
$VH \rightarrow \ell\ell\ell + X  $                                        & 9.7 &100$-$200 & 11.1 & 19.3 & \cite{dzlll} \\
\end{tabular}
\end{ruledtabular}
\end{table*}
\end{center}

\begin{figure}[htb]
\begin{centering}
\includegraphics[width=0.43\textwidth]{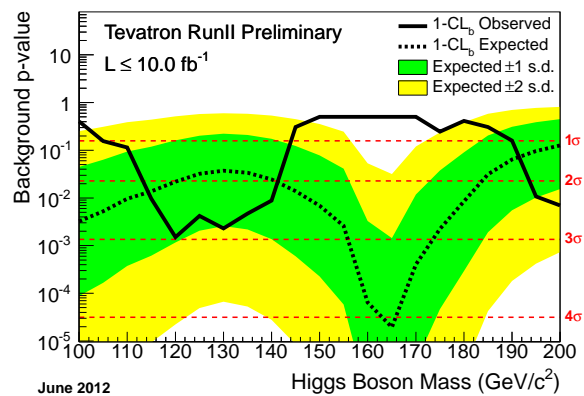} \\
\caption{
\label{comboCLB}
The background $p$ values 1-${\rm C.L.}_{\rm b}$ as a function of the Higgs boson mass (in steps 
of 5 GeV), for the combination of the CDF and D0 analyses. The green and yellow bands 
correspond, respectively, to the regions enclosing 1~$\sigma$ and 2~$\sigma$ fluctuations around the median prediction
in the signal-plus-background hypothesis at each value of $m_H$.}
\end{centering}
\end{figure}

Fig.~\ref{comboCLB} shows 
the $p$ value 1-${\rm CL}_{\rm b}$ as a function of $m_H$, 
i.e., the 
probability that an upward fluctuation of the 
background can give an outcome as signal-like as the data or more.
In the absence of signal, the $p$ value is expected to be uniformly distributed between 0 and 1.
A small $p$ value indicates that the data are unlikely to be explained by the background-only hypothesis.
The smallest observed $p$ value corresponds to a Higgs boson mass of 120 GeV and has
a local significance of 3.0~$\sigma$.
The fluctuations seen in the observed $p$ value as a function of the tested $m_H$ result from excesses seen 
in different search channels, as well as from point-to-point fluctuations 
originating from the separate discriminants used 
at each $m_H$, as  next discussed in more detail. 
The width of the dip from 115 to 
135 GeV is consistent with the combined resolution of the $H \to b\bar{b}$ and $H \to W^+W^-$ channels.   The
effective resolution of this search comes from two independent sources.: the reconstructed candidate
masses, which directly constrain $m_H$, and the expected cross sections times the relevant branching ratios for the
$H \to b\bar{b}$ and $H \to W^+W^-$ channels, which are functions of $m_H$ in the SM.  The observed excess in
the $H \to b\bar{b}$ channels coupled with a less signal-like outcome in the $H \to W^+W^-$ channels
determines the shape of the observed $p$ value as a function of $m_H$.

The strongest sensitivity at low mass comes from the $H \to b\bar{b}$ channels. The largest local significance
in the combination of $H \to b\bar{b}$ channels is 3.3~$\sigma$ at a mass of 135~GeV, while it is
2.8~$\sigma$ at 125~GeV~\cite{comb-diboson}.
\begin{figure}[htb]
\begin{centering}
\includegraphics[width=0.44\textwidth]{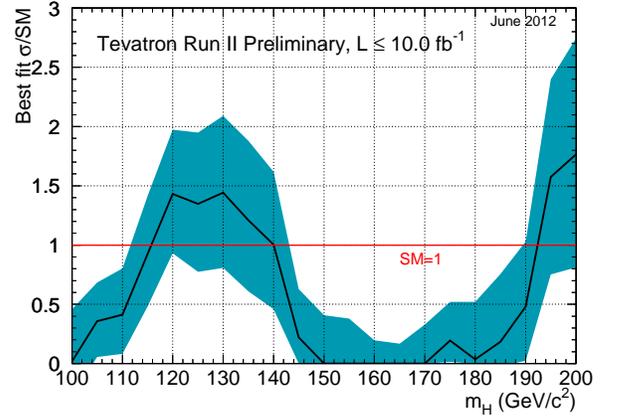}
\caption{
\label{Fit}
The best-fit signal cross section of all CDF and D0 search channels combined shown as a ratio to the 
SM cross section as a function of the tested Higgs boson mass.  The horizontal line at 1 represents the signal 
strength expected for a SM Higgs boson hypothesis.  The blue band shows the 1~$\sigma$ 
uncertainty on the signal fit. 
}
\end{centering}
\end{figure}

In Fig.~\ref{Fit},  the signal strength is allowed to vary as a
function of $m_H$ in the fit of the signal-plus-background hypothesis
to the observed data over the full mass range.
As shown, the resulting best-fit signal strength 
normalized to the SM prediction  
is within 1~$\sigma$ of the SM expectation for a Higgs boson signal in the range 
$110<m_H<140$~GeV.   
The largest signal fit in this range, normalized to the SM prediction, is obtained at 130 GeV,
rather than 
for the smallest $p$-value mass of 120 GeV, since the similar excesses for these
two mass hypotheses
translate into a higher signal strength at 130 GeV.
%
%
The excess in signal-strength around 200~GeV occurs in
a region of low expected sensitivity ($\sim$1~$\sigma$) and with
an unphysically narrow mass range; thus it cannot be
attributed to a SM Higgs boson signal at high mass.

  
At the Tevatron the look-elsewhere effect (LEE) is estimated in a simplified and conservative manner.  
In the mass range 
115--150~GeV, where the low-mass $H\rightarrow b{\bar{b}}$ searches dominate, the reconstructed 
mass resolution is approximately 15\%.  A LEE factor 
of $\simeq 2$ is thus estimated for the low-mass region.  The $H\rightarrow\gamma\gamma$ searches have a much better mass 
resolution, of the order of 3\%, but their contribution to the final LLR is small due to the much smaller 
signal-to-background ratio in those searches.  
The $H\rightarrow \tau^+\tau^-$ searches have both worse reconstructed mass resolution and 
lower signal to background ratio than the $H\rightarrow b{\bar{b}}$ searches, and therefore similarly do not play a significant 
role in the estimation of the LEE. 
The $H \rightarrow WW$ channel has the poorest mass resolution and therefore contributes weakly to the LEE.
 For the combined search of all Tevatron channels, with a conservative LEE of $\simeq$ 4 to take into
account the full range over which the search was performed, 
the global significance of the excess observed at low mass is 
approximately 2.5~$\sigma$

\begin{figure}[htb]
\includegraphics[width=0.47\textwidth]{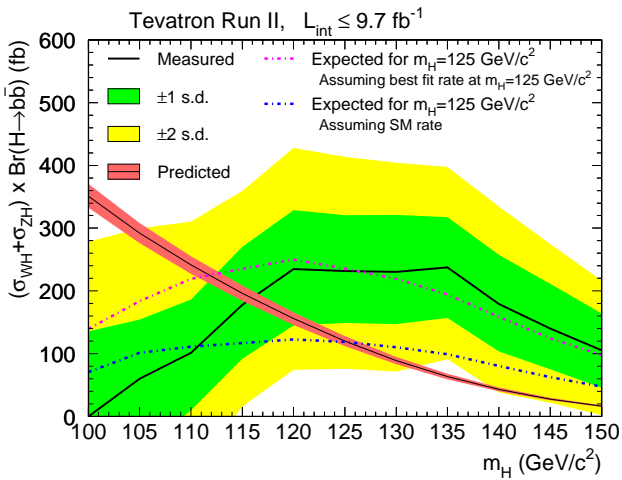} 
\caption{
The best-fit cross section times branching ratio
$\sigma_{WH}+\sigma_{ZH} \times {\cal{B}}(H \rightarrow b \bar{b})$ as a function of $m_H$
measured at the Tevatron.
}
\label{tevbb}
\end{figure}

Applying the low-mass LEE 
to the most significant local $p$ value obtained  
from the CDF+D0 $H\rightarrow b{\bar{b}}$ combination, a global significance of approximately 3.1~$\sigma$ is
obtained, resulting in evidence for the production of a resonance in the $b$-flavored dijet mass distribution,
produced in association with a massive vector boson. Given the mass resolution in this final state, 
this resonance
is consistent with the new boson observed by the LHC experiments, 
and provides the first evidence for fermionic
decays of this boson.

\begin{figure}[hbt]
\includegraphics[width=0.47\textwidth,height=0.28\textheight]{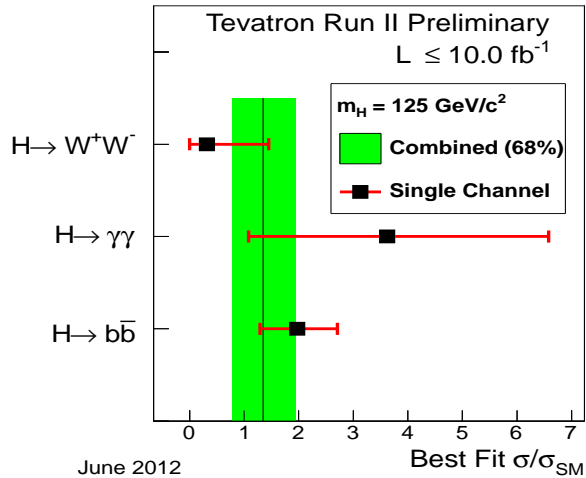} 
\caption{
Best-fit signal strength for the three most sensitive  boson decay modes at the
Tevatron, for $m_H=125$~GeV/$c^2$. The shaded band corresponds to the $\pm$ 1$\sigma$ uncertainty on the full combination.
}
\label{tevsmxsfit}
\end{figure}

The measured cross section times branching ratio
$\sigma_{WH}+\sigma_{ZH} \times {\cal{B}}(H \rightarrow b \bar{b})$ 
is shown in Fig.~\ref{tevbb} as a function of $m_H$.
The resulting value is $0.23^{+0.09}_{-0.08}$ pb for $m_H$ = 125~GeV, consistent with 
the corresponding SM prediction of
0.12 $\pm$ 0.01 pb.
The best fit signal cross section from the combined CDF and D0 
analyses separated into the different Higgs boson decay channels
is shown in Fig.~\ref{tevsmxsfit}, assuming $m_H=125$~GeV.

In summary, at the Tevatron, when combining all search channels,  there is significant excess of data events with respect 
to the background estimation in the mass range $115<m_{H}<135$~GeV.
The $p$ value for a background fluctuation to produce such an excess 
corresponds to a local significance of 3.0 $\sigma$ at 120~GeV.  
The largest excess is observed in the $H\to b\bar{b}$ channels, 
with a local significance of 3.3~$\sigma$,
which results in a global significance of $\approx 3.1$~$\sigma$  when
accounting for look-elsewhere effects.
The CDF and D0 Collaborations thus report evidence for the production of a 
resonance in the $b$-flavored dijet mass distribution produced in association with a massive vector boson,
consistent with the new boson observed by the LHC Collaborations. The measured cross section for this 
process is consistent with the cross section expected for a SM Higgs boson of 125 GeV produced in association
with a $W$ or a $Z$ boson.

	\subsection{Conclusion and prospects}

The LHC experiments have discovered a new boson with mass around 125 GeV and have evidence
for this particle in several decay modes.
The Tevatron experiments report
evidence for a particle, produced in association with $W$ or $Z$ bosons and which decays to $b\bar{b}$, with
a mass compatible to that reported by the LHC experiments.  
The production and decay modes that have been observed
indicate that this boson plays a role in the mechanism of electroweak symmetry breaking and also in
the mass generation for the quarks.  The properties of this boson are compatible with those expected
for a SM Higgs boson but more study is required to fully explore the nature of this discovery.
The discovery of a new boson with properties indicating that it plays a role in electroweak
symmetry breaking is a major breakthrough in fundamental physics.

The LHC experiments expect to integrate up to 30~fb$^{-1}$ of data at 8~TeV center of mass
energy by late 2012.  These data should be sufficient to make first measurements of all accessible
parameters of the boson assuming SM-like behavior.  After the 2012 run the LHC is expected to undergo
a long shutdown to upgrade the energy and luminosity capabilities of the accelerator, to near the design
parameters of 14~TeV and 100~fb$^{-1}$ per year.  Data taken after
the upgrades  should
allow for precision measurements of the boson's properties and exploration of 
non SM physics associated with the boson.

{\it After completion of this review, additional data taken by the ATLAS and CMS collaborations
confirmed that the new discovered boson is a Higgs boson, while the current precision of its
measured properties does not yet allow one to definitely identify it as the standard model Higgs boson.}


\end{document}